%% file: main.tex
\documentclass[aps,prd,reprint,groupedaddress,nofootinbib,superscriptaddress]{revtex4-2}

\usepackage[T1]{fontenc}
\usepackage{amsmath,amssymb,bm,mathtools}
\usepackage{booktabs}
\usepackage{xcolor}
\usepackage[hidelinks]{hyperref}
\usepackage{cleveref}
\usepackage{graphicx}
\usepackage{tikz}
\usepackage[normalem]{ulem}
\usepackage{comment}
\usepackage{subcaption}
\usepackage{float}

\usetikzlibrary{arrows.meta,decorations.pathmorphing,positioning}

\input{tikz_styles.tex}

\newcommand{\dd}{\mathrm d}
\newcommand{\ii}{\mathrm i}
\newcommand{\avg}[1]{\left\langle #1\right\rangle}
\newcommand{\FE}{\mathcal F_E}
\newcommand{\rhoB}{\rho_{\mathrm{bin}}}
\newcommand{\Sorb}{\mathcal S}
\newcommand{\intp}{\int_{\bm p}}
\newcommand{\intk}{\int_{\bm k}}

\begin{document}

\title{Dynamical Friction in an Ultralight Scalar Medium across Coherent and Stochastic Regimes}

\author{Soumodeep Mitra}
\email{soumodeep.mitra@coyotes.usd.edu}
\affiliation{Department of Physics, University of South Dakota, Vermillion, SD 57069, USA}

\author{Jordan Wilson-Gerow}
\email{jwilsong@andrew.cmu.edu}
\affiliation{Department of Physics, Carnegie Mellon University, Pittsburgh, Pennsylvania 15213, USA}

\date{\today}

\begin{abstract}
We study the conservative and dissipative forces on a Newtonian binary interacting with a nonrelativistic ultralight scalar medium. We derive integral expressions for the instantaneous force and the orbit-averaged energy flux generated by wake perturbations of a general background field. For a homogeneous coherent background, the conservative force suffers from a well-known infrared divergence. We trace this divergence to the failure of perturbation theory about a constant scalar state and identify the gravitational Bohr scale at which the homogeneous approximation breaks down. In the long-wavelength regime, the leading dissipation into the medium is quadrupolar, and we obtain an explicit expression for the leading energy flux from an eccentric binary. In the short-wavelength regime, we identify a local ``hard'' region in momentum space that produces a universal Coulomb logarithm, while the nonlogarithmic contribution remains orbit dependent. We then promote the background to a stochastic ensemble with a general velocity distribution. Three independent length scales, together with a derived geometric-mean scale, produce eight distinct scale hierarchies, for which we characterize the ensemble-mean response. Finally, we derive the two-point correlation function of the orbit-averaged energy flux and evaluate it in ``coherent-response'' regimes. Wave interference produces order-unity density fluctuations, so the flux in an individual realization can differ substantially from its ensemble mean.
\end{abstract}

\maketitle

% ==================================================================
\section{Introduction}
\label{sec:introduction}

The observation of gravitational waves emitted from the merger of compact binaries is one of the most potent windows to probe the universe. Current generation detectors from the LIGO-Virgo-KAGRA~\cite{LIGOScientific:2016aoc,LIGOScientific:2020ibl,LIGOScientific:2018mvr,LIGOScientific:2021djp} collaboration have already provided direct experimental probes of General Relativity, black hole physics, and astrophysics~\cite{Cardoso:2019rvt,Liebling:2012fv,Cardoso:2016oxy,Chakravarti:2024ncc,Sarkar:2023rhp,Barausse:2014tra,Speeney:2022ryg,Kavanagh:2020cfn,Speri:2022upm,Cardoso:2021wlq,Cardoso:2022whc,Chakravarti:2025xaj}. With the advent of next-generation detectors e.g. LISA \cite{LISA:2017pwj, LISA:2022kgy,LISA:2022yao}, Einstein Telescope \cite{ET:2019dnz,Punturo:2010zz,Abac:2025saz}, and Cosmic Explorer \cite{Reitze:2019iox}, our sensitivity to strong gravity regimes is expected to reach an unprecedented level. One of the main observational targets of the next-generation detectors are extreme mass-ratio inspirals (EMRIs), where a secondary of small mass $\sim O(1 -10) M_{\odot}$ orbits around a much larger primary $\sim O(10^5-10^7) M_{\odot}$. EMRIs are typically formed  around supermassive black holes found at the center of galaxies, where dense astrophysical environments are expected to exist. 

A binary embedded in such an environment continually reshapes its surroundings. As the bodies orbit, their gravitational fields displace the ambient matter, generating disturbances that subsequently act back on the binary. In its rudimentary form one can imagine a single perturber moving through a material medium, generating a gravitational wake - a local overdensity trailing the perturber - which in turn exerts a gravitational pull back on the perturber. This effect was first studied by Chandrasekhar \cite{chandrasekhar1943dynamical} for an isolated star moving through stellar medium, and is known generally as `Dynamical Friction' (DF).  

For steady straight-line motion, this effect can be summarized by a local drag law. A bound orbit is more subtle. The bodies continually turn through the environment, their wakes can overlap, and the force at one orbital phase can retain memory of disturbances generated at earlier phases. Depending on the scales involved, the medium may resolve the two bodies independently or only through their collective multipole moments. Dynamical friction on a bound orbit is therefore controlled by the relationship between the orbital size and period and the length and time scales over which disturbances propagate.

For a matter-wave medium, the background itself adds another layer to this problem. A spatially constant field is a highly special state. A generic excited field is instead a superposition of waves whose interference produces a time-dependent granular density profile. Depending on the length and time scales being probed, the same background can appear utterly quiescent or as a roiling sea with order-unity density fluctuations. A binary samples this interference pattern along a closed trajectory, while the wake it induces retains some memory of the background encountered at earlier times. Consequently, even binaries with identical orbital parameters and the same ensemble-mean ambient density need not experience the same force.

This observation separates two questions that coincide in a perfectly coherent background. The ensemble-mean force determines the average orbital drift, but it does not determine how representative that drift is for an individual system. An orbit that passes through many independent interference patches may self-average, whereas a binary confined to a single coherent patch can experience a realization-dependent correction of order unity. Which behavior occurs depends jointly on the orbital scales, the scalar dispersion relation, and the spatial and temporal coherence of the background.

Of the various astrophysical media dark matter (DM) holds special interest. After first being theorized following galactic rotation curve observation \cite{rubin1970rotation,rubin1970rotation1} (and various other later observations \cite{SDSS:2005sxd,Clowe:2006eq}), and decades of terrestrial experiments \cite{SuperCDMS:2017mbc,Zatschler:2024ssq, LUX:2018akb}, its nature and properties are still elusive. Galactic and cosmological scale observational data, together with numerical simulations, provide us with a phenomenological \textit{cold dark matter} (CDM) model, which includes details such as its galactic scale distribution \cite{Navarro:1995iw,Navarro:1996gj,Hernquist:1990be}.  However, uncertainty on the behavior of DM at sub-galactic scales and discrepancies between the CDM distribution and observations (known as the `Core-Cusp' problem \cite{de2010core,de_Blok_2009}) of dwarf galaxies \cite{Moore:1994yx,Flores:1994gz,Marchesini:2002vm}  have led to various alternative proposals for the nature of dark matter.

Ultralight fuzzy DM or scalar field dark matter (SFDM) \cite{ferreira2021ultra, Eberhardt:2025caq,hui2017ultralight} is one such exciting proposal. In this model the dark matter is described as an ultralight $\sim(10^{-22}$-$10^{-10})\, \mathrm{eV}$ scalar field which forms a cored density profile  near the center of the galaxy, alleviating the `Core-Cusp' problem.  On the other hand, on length scales larger than the de-Broglie wavelength of the scalars, SFDM reproduces an approximately Navarro-Frenk-White (NFW) distribution and matches with the galactic-scale observation~\cite{Schive:2014dra}. 

The presence of a DM medium near the galactic center is also important due to the formation of overdensity or `spikes' \cite{gondolo1999dark,Sadeghian:2013laa} in the inspiral region, which have been shown to significantly impact gravitational waveforms~\cite{Eda:2013gg, Chakraborty:2024gcr,Rahman:2023sof,speeney2022impact,Mitra:2025tag}, primarily through DF. Such changes in the waveform provide us with a golden opportunity to probe the nature of DM and potentially distinguish between various DM models. 

For wave dark matter, however, the microscopic state underlying such a density profile need not be a single coherent scalar configuration. Virialized wave halos can instead be described by a distribution of occupied modes, whose interference produces spatially and temporally varying density granules~\cite{Schive:2014dra, Lin:2018whl}.  This many-mode description has also been carried through the adiabatic growth of a central black hole: the resulting compressed wave halo is described by a distribution of mode occupations which reduces, in the semiclassical limit, to the familiar particle phase-space distribution~\cite{Kim:2022mdj}. It is therefore natural to ask how DF changes when the scalar background is treated statistically rather than as a perfectly coherent field. 

Significant work has been done to quantify DF in a SFDM medium, including perturbers in straight-line motion through non-interacting scalar fields~\cite{hui2017ultralight,vicente2022dynamical,Mitra:2023sny,Traykova:2023qyv} and superfluids~\cite{Berezhiani:2023vlo,Berezhiani:2025maf}. Buehler and Desjacques studied single and binary perturbers on circular orbits, deriving the force in a coherent background and comparing with numerical calculations in a random-wave background~\cite{Buehler:2022tmr}. Stochastic backgrounds have also been considered for rectilinear motion. Lancaster et al. derived the ensemble-mean drag in an FDM background with finite velocity dispersion, while their random-wave simulations exhibit substantial realization-dependent variation of the instantaneous force~\cite{Lancaster:2019mde}. Separately, the direct gravitational force generated by the pre-existing FDM interference pattern has been studied statistically, including its force correlations, relaxation, and diffusion of embedded objects~\cite{Bar-Or:2018pxz}. These direct stochastic forces are distinct from fluctuations of the wake induced by the perturber itself. What remains missing is an analytic treatment of the realization-dependent induced wake for a general bound binary, including its ensemble mean and fluctuations across the different regimes determined by the orbital and coherence scales.

In this work, we develop a common framework for a binary system immersed in a SFDM environment. Our approach uses the effective field theory framework of~\cite{Modrek:2026, Modrekiladze:2026drr} which, among other things, formulates DF as a generalized self-force problem (see also \cite{Datta:2026eqj}). We treat the SFDM as a free nonrelativistic complex scalar field, inside of which two point masses follow a prescribed periodic orbit. Our object of interest is the gravitational force generated by the wake induced by the binary. This should be distinguished from the direct gravitational force exerted by pre-existing interference granules in the scalar background. We denote the instantaneous positive energy flux into the scalar medium by $\FE(t)$, its average over one orbit in a fixed realization by $\overline{\FE}$, and an ensemble average by angled brackets, $\avg{\mathcal{F}_{E}}$.

We begin in \cref{sec:conservative,sec:small-lambda,sec:large-Lambda} with a homogeneous coherent scalar, which provides the baseline for the stochastic calculation. Before turning to dissipation, \cref{sec:conservative} revisits the infrared divergence in the conservative force generated by the static mass monopole. We show that this divergence signals the failure of perturbation theory about a spatially constant scalar state and identify the gravitational Bohr scale $a_G$ at which this expansion breaks down. The resolution requires a globally consistent, Coulomb-distorted scalar profile.

The orbit-induced dissipative force is well behaved in the long-wavelength regime. When the scalar wavelength is large compared with the orbit, the medium resolves the binary as a composite source. The mass dipole vanishes in the center-of-mass frame, so the leading energy transfer begins at quadrupole order. In \cref{sec:small-lambda}, we derive this contribution and obtain an explicit expression for the leading flux from an eccentric Keplerian orbit.

The opposite limit is considered in \cref{sec:large-Lambda}. When the scalar wavelength is short compared with the orbit, a parametrically broad region of momentum space probes short, locally straight segments of each worldline. We show that this region generates a universal Coulomb logarithm for a general bound trajectory. Both endpoints of the logarithmic interval follow from the calculation: orbital curvature determines its infrared (IR) endpoint, while the scalar dispersion relation and on-shell kinematics determine its ultraviolet (UV) endpoint. The logarithm is local and insensitive to the global orbit, but the accompanying nonlogarithmic terms remain sensitive to the full trajectory.

We promote the coherent background to a stationary stochastic ensemble in \cref{sec:stochastic-mean}. The scalar state is characterized by a general velocity distribution, which introduces spatial and temporal coherence scales in addition to the orbital scale and the scalar de Broglie wavelength. Three independent length scales, together with a derived geometric-mean scale, divide the problem into eight parametrically distinct hierarchies. These regimes distinguish whether the binary acts as a multipolar or locally resolved source and whether the environmental Green's function reduces to its coherent limit or retains the full velocity distribution. We derive the ensemble-mean flux and characterize its leading behavior in all eight cases.

The ensemble mean does not, by itself, determine the force acting in a particular realization. In \cref{sec:stochastic-autocorrelator}, we derive the connected two-time correlation function of the orbit-averaged flux and evaluate it in the coherent regimes. A binary contained within one quasi-static interference patch inherits the order-unity density fluctuations of that patch and does not self-average over a single orbit. By contrast, the universal leading-logarithmic contribution self-averages when the bodies traverse many spatial patches, with the expected inverse-square-root suppression in the number of patches sampled. For the regimes where a kinetic-theory description is more appropriate than a wave-like treatment, we give the exact kernel that governs the fluctuations, while leaving its explicit evaluation to future work.

\begin{figure*}[t]
\centering

\(\displaystyle
\bm F_A(t)
=
\sum_{B=1}^{2}
\left[
%
  % diagram 1
%%%%%%%%%%%%%%%%%%%%%%%%%%%%%%%%%%%%%%%%%%%%%%%%%%%%%%%%%%%%%%%%%%
\begin{tikzpicture}[
    baseline=(current bounding box.center),
    x=0.95cm,
    y=0.95cm
]
  % bottom worldline vertices
  \node[vertexdot] (src) at (-1.9,0) {};
  \node[vertexdot] (rec) at (1.9,0) {};

  \draw[worldline] (-2.9,0) -- (-0.9,0);
  \draw[worldline] (0.9,0) -- (rec);
  \draw[extleg] (rec) -- (2.9,0);

  \node[below=3pt] at (src) {$B$};
  \node[below=3pt] at (rec) {$A$};
  %\node[above=2pt] at (2.75,0) {$i$};

  % upper mixing vertices
  %\node[vertexdot] (mixL) at (-1.15,1.35) {};
  %\node[vertexdot] (mixR) at ( 1.15,1.35) {};

  \node[vertexdot] (mixL) at (-0.95,1.30) {};
  \node[vertexdot] (mixR) at ( 0.95,1.30) {};

  % gravitons
  \draw[graviton] (src) -- (mixL);
  \draw[graviton] (mixR) -- (rec);

  % momentum arrows on gravitons
  \draw[->] (-2.25,0.35) -- (-1.65,1.05);
  \node[left=1pt] at (-1.93,0.72) {$p$};

  \draw[->] (1.65,1.05) -- (2.25,0.35);
  \node[right=1pt] at (1.93,0.72) {$q$};

  % dotted background insertions
  \draw[bgscalar] (-2.15,2.15) -- (mixL);
  \draw[bgscalar] ( 2.15,2.15) -- (mixR);

  \node[above left=1pt]  at (-1.80,1.48) {$k-p$};
  \node[above right=1pt] at ( 1.80,1.48) {$q-k$};

  % internal scalar line across the top
  \draw[scalarflow] (mixL) -- (mixR);
  \node[above=3pt] at (0,1.5) {$k$};
    % momentum arrow on scalar line
  \draw[->] (-0.5, 1.6) -- (0.5,1.60);
\end{tikzpicture}
  \;+\;
  % diagram 2
%%%%%%%%%%%%%%%%%%%%%%%%%%%%%%%%%%%%%%%%%%%%%%%%%%%%%%%%%%%%%%%%%%
\begin{tikzpicture}[
    baseline=(current bounding box.center),
    x=0.95cm,
    y=0.95cm
]
  % bottom worldline vertices
  \node[vertexdot] (src) at (-1.9,0) {};
  \node[vertexdot] (rec) at (1.9,0) {};

  \draw[worldline] (-2.9,0) -- (-0.9,0);
  \draw[worldline] (0.9,0) -- (rec);
  \draw[extleg] (rec) -- (2.9,0);

  \node[below=3pt] at (src) {$B$};
  \node[below=3pt] at (rec) {$A$};
  %\node[above=2pt] at (2.75,0) {$i$};

  % upper mixing vertices
  %\node[vertexdot] (mixL) at (-1.15,1.35) {};
  %\node[vertexdot] (mixR) at ( 1.15,1.35) {};

  \node[vertexdot] (mixL) at (-0.95,1.30) {};
  \node[vertexdot] (mixR) at ( 0.95,1.30) {};

  % gravitons
  \draw[graviton] (src) -- (mixL);
  \draw[graviton] (mixR) -- (rec);

  % momentum arrows on gravitons
  \draw[->] (-2.25,0.35) -- (-1.65,1.05);
  \node[left=1pt] at (-1.93,0.72) {$p$};

  \draw[->] (1.65,1.05) -- (2.25,0.35);
  \node[right=1pt] at (1.93,0.72) {$q$};

  % dotted background insertions
  \draw[bgscalar] (-2.15,2.15) -- (mixL);
  \draw[bgscalar] ( 2.15,2.15) -- (mixR);

  \node[above left=1pt]  at (-1.80,1.48) {$k-p$};
  \node[above right=1pt] at ( 1.80,1.48) {$q-k$};

  % internal scalar line across the top, reversed flow
  \draw[scalarflowrev] (mixL) -- (mixR);
  \node[above=3pt] at (0,1.50) {$k$};

  % momentum arrow on scalar line
  \draw[->] (-0.5, 1.6) -- (0.5,1.60);
\end{tikzpicture}
\right]
\)

\caption{Diagrammatic representation of the leading induced force on body \(A\). Body \(B\) sources a Newtonian perturbation, which excites the scalar background and propagates causally before sourcing the gravitational field felt by body \(A\). The two terms correspond to the two possible complex-scalar orientations. Dashed horizontal lines denote background worldlines, dotted lines denote background scalar insertions, solid scalar lines denote propagating scalar perturbations, and their arrows denote scalar particle-number flow. Separate arrows indicate momentum routing. The explicit Feynman rules and derivation are given in Appendix~\ref{app:feynmandiagrams}.}
\label{fig:master-response}
\end{figure*}
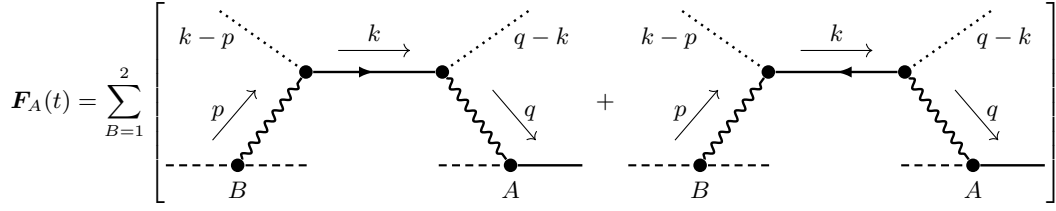

At leading order, all of these effects are organized by the two Feynman diagrams shown in \cref{fig:master-response}. Body $B$ sources a Newtonian perturbation of the scalar background, the resulting scalar disturbance propagates, and its induced gravitational field acts on body $A$. The two diagrams represent the two possible orientations of complex scalar's particle-number flow. The general expressions for the instantaneous force and orbit-averaged flux are derived in Appendix~\ref{app:feynmandiagrams}. The diagrammatic construction follows the framework developed more fully in Ref.~\cite{Modrek:2026}; rather than repeating that formal machinery, we collect the explicit rules and master formulas in the appendix and focus the main text on the scalar-specific scales, limits, and physical consequences.

The sum over $B$ in~\cref{fig:master-response} ensures that both constituents source the environmental wake. This is essential in the long-wavelength regime, where the scalar resolves the binary as a whole and the wakes generated by the two bodies interfere.

% ==================================================================

\section{Conservative force in a coherent scalar background and the infrared problem}
\label{sec:conservative}

We work in the center-of-mass frame,
\begin{equation}
M=m_1+m_2,
\qquad
\mu_{\rm red}=\frac{m_1m_2}{M},
\qquad
\bm r=\bm x_1-\bm x_2,
\label{eq:binary-definitions}
\end{equation}
with
\begin{equation}
\bm x_1=\frac{m_2}{M}\bm r,
\qquad
\bm x_2=-\frac{m_1}{M}\bm r.
\label{eq:cm-worldlines}
\end{equation}
For the prescribed Newtonian Kepler orbit,
\begin{equation}
\ddot{\bm r}=-\frac{GM}{r^3}\bm r,
\qquad
\Omega^2a^3=GM,
\label{eq:relative-kepler}
\end{equation}
where $a$ is the semimajor axis and $T=2\pi/\Omega$ is the orbital period. 

\subsection{Static monopole contribution}
\label{subsec:ir-divergence}

For a homogeneous coherent scalar, the force on body $A$ is derived in Appendix~\ref{app:feynmandiagrams}, and we quote it here. Let $p = |\bm p|$ and $E_{\bm p} = p^{2}/2\mu$. The force is then
\begin{align}
&F_A^i(t) =-\ii(4\pi G)^2\rho_0m_A
\sum_{n=-\infty}^{\infty}
\nonumber \\ 
&\times\intp
 e^{-\ii n\Omega t+\ii\bm p\cdot\bm x_A(t)}
\Sorb_n(\bm p)\frac{p^i}{p^2}\frac{1}{(n\Omega+\ii0)^2-E_{\bm p}^2}\,,
\end{align}
where we have used the periodicity of the source to decompose it into discrete Fourier harmonics,\footnote{The utility of a harmonic-multipole expansion for this problem was demonstrated by~\cite{Eytan:2026kkp}.}
\begin{equation}
\Sorb_n(\bm p)
=\frac{1}{T}\int_0^T\dd t\,
 e^{\ii n\Omega t}
\left[
 m_1e^{-\ii\bm p\cdot\bm x_1(t)}
 +m_2e^{-\ii\bm p\cdot\bm x_2(t)}
\right]\,.
\end{equation}
Center-of-mass conservation implies that its mass dipole vanishes identically.

The infrared-sensitive term comes from the static harmonic.  At small momentum,
\begin{equation}
\Sorb_0(\bm p)=M+\mathcal O(p^2),
\label{eq:static-source-small-p}
\end{equation}
where the linear term vanishes in the center-of-mass frame.  The $n=0$ contribution to the force is
\begin{align}
F^i_{A,0}(t)
&=4\ii(4\pi G)^2\rho_0\mu^2m_AM
\intp e^{\ii\bm p\cdot\bm x_A(t)}\frac{p^i}{p^6}
+\cdots
\notag\\
&=-4(4\pi G)^2\rho_0\mu^2m_AM x_A^j(t)
\intp\frac{p^ip_j}{p^6}+\cdots .
\label{eq:binary-ir-force-integral}
\end{align}
Introducing an infrared momentum cutoff $p_{\min}$,
\begin{equation}
\intp\frac{p^ip^j}{p^6}
=\frac{\delta^{ij}}{6\pi^2p_{\min}},
\label{eq:ir-tensor-integral}
\end{equation}
so that
\begin{equation}\label{eq:binary-ir-force}
F^i_{A,\rm IR}(t)
=-\frac{32}{3}\frac{G^2\rho_0\mu^2}{p_{\min}}\,m_AM\,x_A^i(t).
\end{equation}
This is a conservative, radially directed contribution.  Its dependence on $p_{\min}$ is not a short-distance ambiguity of the binary; it signals that the assumed homogeneous scalar background is not a self-consistent global state in the presence of the static mass monopole.

\subsection{Static scalar distortion and the gravitational Bohr scale}
\label{subsec:monopole-background}

The same conclusion follows directly from computing the scalar perturbation. For a static monopole of mass $M$ and a constant real background $\bar\psi$, the linearized Schr\"odinger equation gives
\begin{equation}
\delta\psi(\omega,\bm p)
=8\pi GM\mu^2\bar\psi\,
(2\pi)\delta(\omega)\frac{1}{p^4}.
\label{eq:deltapsi-p}
\end{equation}
The position-space expression requires an infrared-divergent inverse Fourier transform, which we regulate with a hard cutoff $p_{\min}$,
\begin{equation}
\frac{\delta\psi(\bm x)}{\bar\psi}
=GM\mu^2
\left(-r+\frac{4}{\pi p_{\min}}\right).
\label{eq:deltapsi-x}
\end{equation}
The cutoff-dependent constant term is the same power-law IR divergence that arose in the conservative force, \cref{eq:binary-ir-force}. This constant term is sensitive to the global boundary conditions, and the divergence is an artifact of our assumption that an infinite homogeneous medium is only weakly perturbed by the mass $M$. The linearly growing term is more important, as it identifies the physical scale at which perturbation theory about a constant field breaks down,
\begin{equation}
r=a_G=\frac{1}{GM\mu^2},
\label{eq:bohr-radius}
\end{equation}
which we can identify as the gravitational Bohr radius. 

The breakdown at this scale identifies the physical origin of the infrared divergence. Beyond $r\sim a_G$, the scalar background must be treated nonperturbatively in the central Coulomb potential and develops a Coulomb-distorted profile. Rather than being exactly homogeneous, the appropriate global scalar state is a linear combination of bound and/or scattering Coulomb wavefunctions, depending on the asymptotic boundary conditions. There are no physical solutions that are asymptotically constant. The homogeneous calculation remains a useful local approximation only at radii well inside $a_G$.

\subsection{Physical infrared regulation}
\label{subsec:conservative-regulated}

A consistent conservative calculation must expand around the scalar profile supported by the central gravitational potential and the actual outer boundary conditions. Schematically, the physically correct static radial force on the secondary is of the form
\begin{equation}
\bm F_{2,\rm mono}(r)
=-\frac{Gm_2}{r^2}
\left[m_1+M_{\psi,0}(<r)+\delta M_\psi(<r)\right]\hat{\bm r},
\label{eq:regulated-potential}
\end{equation}
where $M_{\psi,0}(<r)$ is the enclosed mass computed from the unperturbed gravitationally distorted scalar profile and $\delta M_\psi(<r)$ is the additional redistribution sourced by the secondary.  Equation~\eqref{eq:regulated-potential} is finite once the global profile is specified. 

In the homogeneous computation above, the background scalar profile was constant at large distance, in direct conflict with the exponentially decaying or oscillatory behavior of Coulomb wavefunctions. The inconsistency of these global boundary conditions manifested itself as a pathological prediction of a divergent $\delta M_{\psi}(<r)$.

The infrared divergence in the conservative force was identified by Buehler and Desjacques~\cite{Buehler:2022tmr} in their study of circular orbits, and subsequently discussed by Berezhiani et al.~\cite{Berezhiani:2023vlo} and Koo and Lee~\cite{Koo:2025jkx}. Buehler and Desjacques related the divergence to the diffusive Schrödinger response and the finite extent of the medium, while Berezhiani et al. identified the Jeans scale as a natural regulator once self-gravity is restored. Here we identify an additional, parametrically earlier breakdown: in the gravitational field of the central monopole, perturbation theory about a homogeneous scalar state fails at the gravitational Bohr scale $a_G$.

The Jeans length relevant for the free scalar field is the so-called quantum-pressure dominated scale~\cite{Berezhiani:2023vlo},\footnote{Here and throughout the paper, we omit factors of $2\pi$ relating inverse wavenumbers to wavelengths.}
\begin{equation}
    \lambda_{\rm J} = \left(\frac{1}{16\pi G\rho_{0}\mu^2}\right)^{1/4}.
\end{equation}
There is an intuitive way to compare this scale with the gravitational Bohr radius. Let $M_{\psi}(a_{G})$ be the total mass of the homogeneous scalar field that is enclosed in a region of size $a_{G}$, 
\begin{equation}
    M_{\psi}(a_{G})=\frac{4\pi}{3}\rho_0a_{G}^{3}\,.
\end{equation}
The ratio of IR length scales is then
\begin{equation}
    \frac{a_G}{\lambda_{\rm J}} = \left[12\frac{M_{\psi}(a_G)}{M}\right]^{1/4}\,,
\end{equation}
up to order-unity factors. Thus, when the total binary mass dominates the local mass of the environment, the gravitational Bohr radius is parametrically smaller than the Jeans length. It is therefore $a_G$ that determines the effective infrared cutoff $p_{\min}$, not $\lambda_{\rm J}$. This changes the predicted conservative force by orders of magnitude.

\subsection{Separation from the dissipative orbital scales}
\label{subsec:atom-parametrics}

The breakdown of perturbation theory indicates that the physically correct background involves Coulomb wavefunctions rather than a homogeneous field. In this work, and in most other computations in the literature,\footnote{See \cite{Tomaselli:2023ysb} (and references therein) for exceptions that compute dynamical friction in a gravitational atom.} the explicit calculations are performed for a homogeneous field. It is therefore important to ask whether these calculations remain relevant for a Coulomb-distorted background.

Intuitively, one expects the homogeneous approximation to be valid when the orbit lies well within the gravitational atom. This condition can be stated neatly in terms of the scalar field's de Broglie wavelength.

For a Keplerian orbit of characteristic radius $R$ and speed $v$, $v^2\sim GM/R$.  With the de Broglie wavelength defined as
\begin{equation}
\lambda_{\rm dB}\equiv\frac{1}{\mu v},
\label{eq:lambda-db-definition}
\end{equation}
we find
\begin{equation}
\frac{R}{a_G}
=RGM\mu^2
\sim\left(\frac{R}{\lambda_{\rm dB}}\right)^2.
\label{eq:R-over-aG}
\end{equation}
Relatedly, the ratio of gravitational-atom level spacing to the orbital frequency $\Omega\sim v/R$ scales as
\begin{equation}
\frac{\Delta E_{\rm atom}}{\Omega}
\sim\left(\frac{R}{\lambda_{\rm dB}}\right)^3.
\label{eq:level-spacing-over-omega}
\end{equation}
Thus, when $R\ll\lambda_{\rm dB}$, the orbit can lie well inside the core of the Coulomb wavefunction, parametrically inside the scale at which the homogeneous background assumption fails. In this same limit the orbital frequency remains large compared with the gravitational-atom level spacing so the discreteness of the bound spectrum should manifest only as a subleading correction. 

We expect that our general results for dissipative forces in homogeneous coherent fields, computed in subsequent sections, remain valid at leading order in small $R/\lambda_{\rm dB}$, despite the IR divergence  encountered in the conservative sector. For generic values of $R/\lambda_{\rm dB}$ this will no longer be the case and one must instead perform computations with the globally correct background field configuration.

% ==================================================================
\section{Long-wavelength coherent dissipation from an isolated binary}
\label{sec:small-lambda}

The positive orbit-averaged energy flux in a constant coherent background is
\begin{equation}
\overline{\FE}
=\pi(4\pi G)^2\rho_0
\sum_{n=1}^{\infty}\intp
\frac{|\Sorb_n(\bm p)|^2}{p^2}
\delta\!\left(n\Omega-E_{\bm p}\right),
\label{eq:binary-coherent-flux-main}
\end{equation}
which is Eq.~\eqref{eq:coherentfluxformula} of Appendix~\ref{app:feynmandiagrams}. This captures the ``friction'' part of the force. It includes the $m_1^2$, $m_2^2$, and $m_1m_2$ interference terms and does not generically reduce to a sum of two independent one-body drag forces. In the long-wavelength regime, $R/\lambda_{\rm dB}\ll1$, the scalar resolves the entire binary orbit and interference terms proportional to $m_1m_2$ are relevant. By contrast, in the short-wavelength limit considered in \cref{sec:large-Lambda}, interference effects are suppressed and the leading logarithmic flux reduces to a sum of independent one-body contributions.

The magnitude of the on-shell momentum at harmonic $n$ is
\begin{equation}
p_n=\sqrt{2\mu n\Omega}
\Longrightarrow
p_na=\sqrt{n \times 2 \mu v a}  = \sqrt{n\zeta}
\label{eq:on-shell-p}
\end{equation}
where we have defined, $\zeta \equiv2\mu\Omega a^2 = 2\mu va$, which is the dimensionless ratio of the orbit size, $a$, to the de Broglie wavelength of the scalar $\lambda_{\rm dB}$. The size of $\zeta$ controls which regime the system is in. The long-wavelength multipole regime for the low harmonics is given by $\zeta\ll1$. In this regime we can multipole expand the source $\Sorb_n(\bm p)$, with each term given by powers of $(p\cdot a)$ with $p$ fixed to be on-shell via the delta function in \eqref{eq:binary-coherent-flux-main}. 

We would like to emphasize that we have not made an ad hoc assumption about $a$ or $\lambda_{\rm dB}$ serving as UV or IR cutoffs in the problem. Rather, the kinematics has selected this ratio of physical scales as the one which controls the physics, and we will be careful to track the explicit dependence on $\zeta$ throughout our computations.\footnote{This will be particularly relevant for~\cref{sec:large-Lambda}.}

\subsection{Multipole expansion and dipole cancellation}
\label{subsec:binary-multipoles}

For $n\neq0$, the binary source has the Cartesian multipole-moment expansion
\begin{align}
\Sorb_n(\bm p)
={}&-\ii p_iD_n^i
-\frac12p_ip_jI_n^{ij}
+\mathcal O((pa)^3),
\label{eq:binary-multipole-expansion}
\\
D_n^i
={}&\frac{1}{T}\int_0^T\dd t\,
\left(m_1x_1^i+m_2x_2^i\right)e^{\ii n\Omega t}=0,
\label{eq:dipole-cancellation}
\\
I_n^{ij}
={}&\frac{1}{T}\int_0^T\dd t\,
\left(m_1x_1^ix_1^j+m_2x_2^ix_2^j\right)e^{\ii n\Omega t}
\notag\\
={}&\frac{\mu_{\rm red}}{T}\int_0^T\dd t\,
r^i(t)r^j(t)e^{\ii n\Omega t}.
\label{eq:binary-second-moment}
\end{align}
Here $I^{ij}$ is the trace-full symmetric second mass moment.  For eccentric motion its trace is time dependent and contributes at the same order as the symmetric trace-free part.

Inserting Eq.~\eqref{eq:binary-multipole-expansion} into Eq.~\eqref{eq:binary-coherent-flux-main}, and using
\begin{equation}
\int\dd\Omega_{\bm p}\,
\hat p_i\hat p_j\hat p_k\hat p_l
=\frac{4\pi}{15}
\left(\delta_{ij}\delta_{kl}
+\delta_{ik}\delta_{jl}
+\delta_{il}\delta_{jk}\right),
\label{eq:fourth-angular-average}
\end{equation}
we obtain
\begin{equation}
\overline{\FE}
=\frac{4\pi G^2\mu_{\rm red}^2\rho_0}{v}
\frac{\zeta^{5/2}}{60}
\left[\mathcal Q(e)+\mathcal O(\zeta)\right]\,,
\label{eq:quadrupole-flux-normalized}
\end{equation}
where
\begin{equation}
\mathcal Q(e)
\equiv
\sum_{n=1}^{\infty}n^{3/2}
\left[
2\widetilde I_n^{ij}\widetilde I_{-n}^{ij}
+\widetilde I_n^{ii}\widetilde I_{-n}^{jj}
\right]\,,
\label{eq:Qe-definition}
\end{equation}
and the dimensionless quadrupole harmonic is
\begin{equation}
\widetilde I_n^{ij}
\equiv\frac{I_n^{ij}}{\mu_{\rm red}a^2}\,.
\label{eq:dimensionless-quadrupole}
\end{equation}

Because the binary mass dipole vanishes, the leading source for the flux is the dynamical quadrupole moment. This cancellation is an interference effect between the $m_1$ and $m_2$ sources. Each mass individually sources a dipolar response and produces a flux proportional to $\zeta^{3/2}$ at leading order. This scaling was also found in an explicit computation for a single source on a circular orbit~\cite{Koo:2025jkx}. For a binary, however, this dipole contribution is canceled by the other constituent, so the leading behavior is quadrupolar and proportional to $\zeta^{5/2}$.

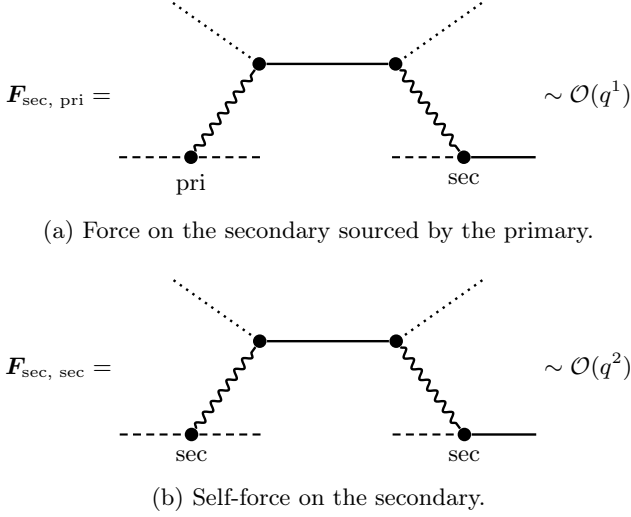
\begin{figure}[h]
\centering
\begingroup

\newcommand{\EMRIdiagram}[2]{%
  \begin{tikzpicture}[
      baseline=(current bounding box.center),
      x=0.95cm,
      y=0.95cm
  ]
    \node[vertexdot] (src) at (-1.9,0) {};
    \node[vertexdot] (rec) at (1.9,0) {};

    \draw[worldline] (-2.9,0) -- (-0.9,0);
    \draw[worldline] (0.9,0) -- (rec);
    \draw[extleg] (rec) -- (2.9,0);

    \node[below=3pt] at (src) {$#1$};
    \node[below=3pt] at (rec) {$#2$};

    \node[vertexdot] (mixL) at (-0.95,1.30) {};
    \node[vertexdot] (mixR) at ( 0.95,1.30) {};

    \draw[graviton] (src) -- (mixL);
    \draw[graviton] (mixR) -- (rec);

    \draw[bgscalar] (-2.15,2.15) -- (mixL);
    \draw[bgscalar] ( 2.15,2.15) -- (mixR);

    \draw[line width=0.9pt] (mixL) -- (mixR);
  \end{tikzpicture}%
}

\begin{subfigure}[b]{\linewidth}
  \centering
  \(
    \displaystyle
    \bm{F}_{\text{sec, pri}} = \EMRIdiagram{\text{pri}}{\text{sec}} \sim\mathcal{O}(q^1)
  \)
  \caption{Force on the secondary sourced by the primary.}
  \label{fig:force-sec-pri}
\end{subfigure}

\vspace{1em}

\begin{subfigure}[b]{\linewidth}
  \centering
  \(
    \displaystyle
    \bm{F}_{\text{sec, sec}} = \EMRIdiagram{\text{sec}}{\text{sec}}\sim\mathcal{O}(q^2)
  \)
  \caption{Self-force on the secondary.}
  \label{fig:force-sec-sec}
\end{subfigure}

\endgroup
\caption{Schematic diagrammatic representation of the force on the secondary body in an EMRI. Momentum flow and scalar orientations are ignored, but otherwise the conventions of \cref{fig:master-response} are used. The sum over the source body $B$ is now expanded into two subfigures.}
\label{fig:EMRI-force}
\end{figure}

To our knowledge this has not been pointed out yet in the literature, and it can be counterintuitive in the case of an extreme mass-ratio system so we would like to restate it more clearly. The force on the secondary body is given by the sum of the two diagrams in \cref{fig:EMRI-force}. To keep the presentation compact we have omitted the momentum labels, and we have omitted drawing both of the complex scalar particle-flow orientations in each subfigure. 

If we denote the mass ratio by $q$, then the naive power counting of the two diagrams is $\bm F_{\rm sec, pri}\sim \mathcal{O}(q^{1})$ and $\bm F_{\rm sec, sec}\sim \mathcal{O}(q^{2})$. It appears as if these diagrams cannot interfere with each other in the $q\ll1$ limit. Moreover, in this limit one intuitively expects the secondary to orbit around a fixed-position primary that only creates a static deformation of the environment and doesn't dissipate energy to a wake.

This intuition is correct, but only at the leading $q^{1}$ order. The above naive power counting only determines the leading contribution from each diagram, but in reality each diagram contains a tower of subleading contributions given by powers of $q$. It is the first sub-leading correction to diagram~(\ref{fig:force-sec-pri}) which interferes with the leading contribution from  diagram~(\ref{fig:force-sec-sec}). From the equations of the Keplerian orbit, \cref{eq:cm-worldlines}, if we take particle $1$ as the primary and particle 2 as the secondary, we see that their positions satisfy
\begin{align}
    \bm x_{\rm sec} &= -\bm r +\mathcal{O}(q^1)\,, \nonumber \\
    \bm x_{\rm pri} & = q\bm r + \mathcal{O}(q^2)\,.
\end{align}
Thus, when we expand the position of the primary to $\mathcal{O}(q^1)$ (dipole order) when computing diagram~(\ref{fig:force-sec-pri}), we obtain exactly the $q\bm r$ contribution necessary for it to cancel the dipole order part of diagram~(\ref{fig:force-sec-sec}).

From the mass multipole-moment expansion
\begin{equation}
    Q^{i_{1}...i_{m}}\sim m_{\rm pri}\big(x^{i_1}_{\rm pri}...x^{i_m}_{\rm pri}+q x^{i_1}_{\rm sec}...x^{i_m}_{\rm sec}\big)
\end{equation}
it is easy to see that in the extreme mass ratio limit the above cancellation only occurs at dipole order, and so one can safely omit the primary when computing the quadrupole (and higher multipole) contributions to the dissipative force.

This feature has a close analogue in the vacuum gravitational self-force problem. In the usual black-hole perturbation theory description of an EMRI, the Schwarzschild background is centered on a fixed primary, so the $\mathcal O(q)$ motion of the primary about the center of mass is not represented by an explicit worldline. Nevertheless, the corresponding recoil information is not absent from the perturbation theory. Detweiler and Poisson showed that it is carried by the nonradiative, even-parity $\ell=1$ metric perturbation, which is associated with the motion of the central black hole about the system's center of mass and contributes to the self-force already at Newtonian order~\cite{Detweiler:2003ci}.

In a point-particle description the same bookkeeping can instead be made explicit by retaining the motion of the primary, as we have done here. Alternatively, in the extreme-mass-ratio EFT the displacement of the heavy worldline can be integrated out, generating a recoil operator which systematically encodes its response to the secondary~\cite{Cheung:2023lnj,Cheung:2024byb}.

Thus, although the primary can be treated as fixed at leading order in the mass ratio, its recoil must be retained at the subleading order relevant for the dipole cancellation.

\subsection{Eccentric Keplerian orbit}
\label{subsec:eccentric-quadrupole}

For eccentric Keplerian orbits we can compute the quadrupole harmonics exactly. Define the mean anomaly $\mathcal M=\Omega t$ and eccentric anomaly $u$ by
\begin{equation}
\mathcal M=u-e\sin u,
\qquad
\dd\mathcal M=(1-e\cos u)\dd u.
\label{eq:kepler-anomaly}
\end{equation}
The relative orbit is parameterized as
\begin{equation}
r^x(u)=a(\cos u-e),
\qquad
r^y(u)=a\sqrt{1-e^2}\sin u\,,
\label{eq:ellipse-cartesian}
\end{equation}
and the dimensionless quadrupole harmonic is then given by the integral
\begin{equation}
\widetilde{I}_n^{ij}
=\frac{1}{2\pi a^{2}}
\int_0^{2\pi}\dd u\,
(1-e\cos u)r^i(u)r^j(u)
 e^{\ii n(u-e\sin u)}.
\label{eq:eccentric-quadrupole-integral}
\end{equation}
The integrals can be performed exactly, and for $n\geq1$ the nonzero components can be written compactly as
\begin{align}
\tilde{I}_n^{xx}
&=\frac{2}{n^2e^2}
\left[-J_n(ne)+ne(1-e^2)J_n'(ne)\right],
\label{eq:Ixx-eccentric}
\\
\tilde{I}_n^{yy}
&=\frac{2(1-e^2)}{n^2e^2}
\left[J_n(ne)-neJ_n'(ne)\right],
\label{eq:Iyy-eccentric}
\\
\tilde{I}_n^{xy}
&=\frac{2\ii \sqrt{1-e^2}}{n}
J_n''(ne).
\label{eq:Ixy-eccentric}
\end{align}

The harmonic sum over Bessel functions gives the exact leading-order flux. It converges within a few terms away from the extreme $e\to1$ limit. More terms are required to cover the full eccentricity range; however, we find a compact polynomial fit
\begin{equation}
\mathcal Q(e)
\simeq
1.42+1.98e^2-0.25e^4-0.41e^6 \,,
\label{eq:eccentric-fit}
\end{equation}
which is accurate to better than $0.5\%$ over $0\leq e<1$. The exact circular limit is
\begin{equation}
\mathcal Q(0)=\sqrt{2}\,.
\end{equation}
The flux is then well described by
\begin{align}
\overline{\FE}
=&\frac{4\pi G^2\mu_{\rm red}^2\rho_0}{v}
\frac{\zeta^{5/2}}{60} \nonumber \\
&\times
\left[1.42+1.98e^2-0.25e^4-0.41e^6+\mathcal O(\zeta)\right]\,,
\label{eq:circular-quadrupole-flux}
\end{align}
which is one of the primary novel results of this work. 

This section was focused on $\zeta \ll1$, large-wavelengths, or equivalently, small-orbits. In the following section we complement this analysis by investigating the opposite limit, $\zeta \gg1$.

% ==================================================================
\section{Short-wavelength dynamical friction and the universal logarithm}
\label{sec:large-Lambda}

The logarithmic contribution to dynamical friction is local and insensitive to the global trajectory. In Ref.~\cite{Modrek:2026}, we showed using renormalization-group methods that the coefficient of this logarithm is universal for arbitrary motion. That argument does not, however, determine the physical scales that delimit the logarithmic region. Here we specialize that analysis to a coherent nonrelativistic scalar and identify the UV and IR endpoints appropriate for a bound orbit. For a  general  trajectory with characteristic size $R$, in the limit  $R\gg\lambda_{\rm dB}$ we will find the logarithmic momentum interval
\begin{equation}
\sqrt{2\mu\Omega}\,\,\ll \,\,p\,\,\ll \,\,2\mu v,
\end{equation}
which produces $\log\sqrt{\zeta}$. The lower endpoint,
$$p_\Omega=\sqrt{2\mu\Omega}\,,$$ marks where the local straight-line approximation breaks down and the curvature of the orbit becomes relevant, while the upper endpoint follows from the scalar dispersion relation and the on-shell kinematics.

We restrict to the wave-regulated regime $\lambda_{\rm dB}\gg R_{90}$, where $R_{90}\sim Gm/v^2$ is the impact parameter below which scattering from the body becomes nonperturbative. In this hierarchy, the wave nature of the matter cuts off the logarithmic region before strong gravitational scattering becomes important.

We start from the instantaneous force on one of the bodies, Eq.~\eqref{eq:forceinstant2general}, specialized to the coherent-field response in Eq.~\eqref{eq:coherent-chi}. Writing the transferred four-momentum as $p=(\omega,\bm p)$, the force becomes
\begin{align}\label{eq:forcelaw_largeLambdasection}
&F_A^i(t) = -\ii(4\pi G)^2 \rho_0 m_A \sum_{B=1}^{2}m_{B}
\int\dd t' \nonumber \\
&\times \int_{p}
 e^{-\ii \omega(t-t')+\ii \bm p \cdot (\bm x_A(t)-\bm x_B(t'))}
\frac{p^i}{p^2}
\frac{1}{(\omega+\ii 0)^2-E_{\bm p}^2}\,.
\end{align}
Throughout this section, $R$ and $v$ denote characteristic orbital scales. Relations involving them are understood parametrically, up to eccentricity-dependent factors of order unity, which affect only the nonlogarithmic part of the force.

We assume in this section that $A$ is the lighter body (the secondary), so the size of its orbit is of the same order as the relative Keplerian orbit, $x_A\sim R$, for all mass ratios. For comparable masses, the following arguments also apply to the force on the primary, and the force on each body contains a large logarithm. If the primary is significantly heavier than the secondary, define the small mass ratio $q\equiv m_{\rm sec}/m_{\rm pri}$. The primary's orbital radius and speed then scale as $q R$ and $q v$, respectively. The force on the secondary contains a large logarithm when $R\gg\lambda_{\rm dB}$, whereas the force on the primary does so only under the stronger condition $q^2R\gg\lambda_{\rm dB}$.

Let $\tau=t-t'$ and $\bm r_{AB}(\tau)=\bm x_A(t)-\bm x_B(t-\tau)$.  For generic orbital separations, $|\bm r_{AB}|\sim R$ and $\tau\sim \Omega^{-1} = R/v$. Intuitively, the natural frequency scale for the environment is $E_{\bm p}$, and the two Fourier phases then scale as
\begin{equation}
    p\Delta x\sim pR\,,\qquad \omega\Delta t\sim \frac{E_{\bm p}}{\Omega}=\frac{p^2}{2\mu\Omega}\,.
\end{equation}
This scaling becomes more precise after integrating over $\omega$, which gives
\begin{align}\label{eq:forcelaw_largeLambdasection2}
&F_A^i(t) = \nonumber \\
&\ii(4\pi G)^2\rho_0 m_A \sum_{B=1}^{2}m_{B}
\int\dd \tau\,\theta(\tau) \int_{\bm p}
 e^{\ii \bm p \cdot \bm r_{AB}(\tau)}\frac{\sin(E_{\bm p}\tau)}{E_{\bm p}}
\frac{p^i}{p^2}\,,
\end{align}
Since $E_{\bm p}=p^2/(2\mu)$, the radial momentum measure is nominally $\dd p/p$. However, the orbital phases determine whether this nominal logarithm is actually realized.

The integral divides naturally into three regions according to the sizes of the Fourier phases: $p<R^{-1}$, $R^{-1}<p<p_\Omega$, and $p>p_\Omega$. The two boundaries answer different physical questions: $p\sim R^{-1}$ determines whether a mode spatially resolves the orbit, while $p\sim p_\Omega$ determines whether the medium responds on a timescale shorter than the orbital period and can go on shell. These boundaries are well separated,
\begin{equation}
    p_{\Omega} = (2\mu\Omega R^2)^{1/2}R^{-1} \sim \sqrt {\zeta}R^{-1} \gg R^{-1}\,. 
\end{equation}

\subsection{Nonlogarithmic response}
\label{subsec:large-soft}

For both of the regions with $p< p_{\Omega}$, we have $E_{\bm p} < \Omega$. As such, a binary with orbital frequency $\Omega$ cannot excite modes of the environment on-shell and there will be no contribution to the orbit-averaged dissipative flux from these regions. This intuitive argument can be made precise using a harmonic decomposition, \cref{eq:coherentfluxformula}, where the flux integral has zero support for $E_{\bm p} <\Omega$. Thus these two regions, soft ($p\lesssim R^{-1}$) and intermediate ($R^{-1}<p<p_\Omega$), \emph{only generate conservative forces}. Furthermore, as we will demonstrate, they only contribute at $\mathcal{O}(\zeta^{0})$ in a large $\zeta$ expansion. Since the static monopole contribution to the force was treated separately in \cref{sec:conservative}, in what follows we focus on the nonstatic, orbit-induced force. 

First consider the soft region, $p\lesssim R^{-1}$. We rescale to dimensionless variables
\begin{align}
\bm p&=R^{-1}\widetilde{\bm p},
\qquad
\omega=\frac{v}{R}\widetilde{\omega}, \nonumber \\
\tau &=\frac{R}{v}s,
\qquad\!
\bm r_{AB}(\tau)=R\widetilde{\bm r}_{AB}(s)\,.
\end{align}
The soft contribution to the force is then
\begin{align}
&F_{A}^{i\,\,\rm soft}
=\frac{-\ii (4\pi G)^2\rho_0}{v^2}m_{A}\sum_{B=1}^2m_{B}\nonumber \\
&\times \int\dd s\int_{|\widetilde{\bm p}|\lesssim 1}\frac{\dd^{4}\widetilde p}{(2\pi)^{4}}e^{-\ii \widetilde{\omega}s+\ii \widetilde{\bm p}\cdot \widetilde{\bm r}_{AB}(s)}\frac{\widetilde{p}^{i}}{\widetilde{\bm p}^2}\frac{1}{(\widetilde{\omega}+\ii 0)^2-\zeta^{-2}\widetilde{\bm p}^4}\,.
\label{eq:soft-region}
\end{align}
After factoring out the relevant dimensionful quantities, we see that the soft contribution is an order-unity functional of the global trajectory. The large parameter $\zeta$ appears only in a term suppressed within the Green's function. The soft force, therefore,  is not necessarily small in the appropriate units, but it does not have a large logarithm.

Next consider the intermediate region, $R^{-1}\leq p<p_\Omega$. The rescaling argument above does not directly apply, but the conclusion is the same. This region is parametrically broad when $\sqrt{\zeta}\gg1$, so one might expect a correspondingly large contribution to the force. This does not occur. Since the modes cannot go on shell for $p<p_\Omega$, there is no resonant enhancement from the pole in the Green's function. Moreover, the spatial phase is rapidly oscillating for $pR\gg1$, causing contributions from the broad interval to cancel.

To see this, consider the $\tau$ integral in \cref{eq:forcelaw_largeLambdasection2}. The spatial phase $\bm p\cdot \bm r_{AB}(\tau)\sim pR\gg 1$ is rapidly oscillating, so a stationary phase approximation is appropriate. The phase is $\Phi(\tau) = \bm p\cdot \bm r_{AB}(\tau)$, and about each stationary point $\tau_s$ we have the gaussian approximation
\begin{equation}
    \Phi(\tau)\approx \Phi(\tau_s)+\frac{1}{2}(\tau-\tau_s)^{2}\bm p\cdot \ddot{\bm {r}}_{AB}(\tau_s)\,.
\end{equation}
Using the characteristic acceleration scale, $\ddot{\bm{r}}\sim R\Omega^2$, the saddle integral yields a factor proportional to $(pR\Omega^2)^{-1/2}$. It follows that in this intermediate region the $|\bm p|$ integral from \cref{eq:forcelaw_largeLambdasection2} has the form
\begin{equation}
    \frac{F^{\rm intermediate}}{F_0} \sim \int^{p_{\Omega}}_{R^{-1}} \frac{\dd p}{p} (pR)^{-1/2} = 2\left(1-\zeta^{-1/4}\right)\,,
\end{equation}
where
\begin{equation}
    F_{0}\sim \frac{(4\pi G)^2\rho_0 m_A M}{v^2}\,.
\end{equation}
 The integral yields no large logarithm, nor positive powers of $\zeta$. Thus, the contribution to the force from this region is also order-unity in the appropriate units.

\subsection{Local response and the universal logarithm}

We now turn to the ``hard'' region, $p>p_\Omega$. Momenta near its lower boundary contribute an order-unity force in the same units as Eq.~\eqref{eq:soft-region}. When $R\gg\lambda_{\rm dB}$, however, the hard region also contains a parametrically broad interval with $p\gg p_\Omega$ that produces the large logarithm.

For $p\gg p_\Omega$, the momentum integral is incoherent at generic time separations. At short time separations, however, $\bm r_{AA}(\tau)$ has not yet grown to the orbital scale $R$, and the integral can remain coherent. In contrast, the separation $\bm r_{AB}(\tau)$ with $B\neq A$ remains of order $R$. Large-momentum modes are therefore incoherent across the entire orbit, so the $B\neq A$ terms do not contribute to the leading logarithm, which is determined only by the local environment of each body.

Expanding in the short time limit the Fourier phase takes the schematic form
\begin{equation}
\bm p \cdot\bm r_{AA}(\tau)
\sim pv\tau + p\frac{v^2}{R}\tau^2
+\cdots\,
\label{eq:short-time-trajectory}
\end{equation}
where we have used the characteristic acceleration $v^2/R$. The phase can remain coherent in the hard region when the energy term $E_{\bm p}\tau$ balances the linear term $pv\tau$ and the higher-order acceleration terms can be neglected. Equation~\eqref{eq:short-time-trajectory} shows that the latter condition holds for
\begin{equation}
    \frac{1}{pv}\ll \tau \ll \sqrt{\frac{R}{pv^2}} = \frac{1}{pv}(pR)^{1/2}\,.
\end{equation}
This domain is parametrically large precisely in the part of the hard region with $pR\gg1$. Furthermore, for $p\gg p_{\Omega}$ we have $E_{\bm p}\gg \Omega$, so the environmental response time is much shorter than an orbital period and the response is local in time and insensitive to the global details of the orbit. Together these conditions justify approximating the orbit by its local tangent,\footnote{Since we only need to worry about one body now, we will keep the subscript $A$ implicit for the rest of the section.}
\begin{align}
&F^{i\,\,\rm LL}(t) = \nonumber \\
&-\ii(4\pi G)^2 \rho_0 m^2
\int\dd \tau  \int_{p}
 e^{-\ii (\omega- \bm p \cdot \bm v(t))\tau}
\frac{p^i}{p^2}
\frac{1}{(\omega+\ii 0)^2-E_{\bm p}^2}\,.
\label{eq:hard-region-integral}
\end{align}
To be clear, this approximation holds for the part of the hard region with $p\gg p_{\Omega}$. For $p\gtrsim p_{\Omega}$ there are still $\mathcal{O}(1)$ contributions to the force, both dissipative and conservative, which are sensitive to the whole orbit. The expression in \cref{eq:hard-region-integral} captures just the leading-log part of the hard region.

Only the imaginary part of the Green's function contributes, producing a delta function that enforces the on-shell condition $\omega=\pm E_{\bm p}$. The $\tau$ integral produces another delta function enforcing $\omega=\bm p\cdot\bm v(t)$. These conditions can be satisfied simultaneously only for $p<2\mu v$. The integral then reduces to
\begin{equation} 
F^{i\,\,\rm LL}(t) = -\FE(t)\frac{v^{i}}{v^2},
\end{equation}
with $v=|\bm v(t)|$ being the instantaneous speed of the body, and with the instantaneous energy flux given by
\begin{align}
&\FE(t) = \frac{4\pi (Gm)^2\rho_0}{v}
\int_{p_{\Omega}}^{2\mu v}\frac{\dd p}{p}\,,
\label{eq:hard-region-integral2}
\end{align}
Hence we can see that the hard region produces the exact universal logarithm \cite{Modrek:2026} of the form $\ln\left(\frac{p_{\rm{UV}}}  {p_{\rm{IR}}}\right)$. The upper momentum cutoff (UV scale) is given by $p_{\rm UV}=2\mu v=2/\lambda_{\rm dB}$, and corresponds to the short-distance scale $b_{\min}=p_{\rm UV}^{-1}=\lambda_{\rm dB}/2$. Because we have assumed $\lambda_{\rm dB}\gg R_{90}$, wave dispersion terminates the logarithmic region before gravitational scattering becomes nonperturbative. In the opposite hierarchy, $\lambda_{\rm dB}\lesssim R_{90}$, the wave calculation must instead be matched onto the short-distance particle-scattering problem, and the ultraviolet length scale entering the Coulomb logarithm is of order $R_{90}$~\cite{Modrek:2026}. On the other hand, the lower momentum cutoff (IR scale) is found to be 
$$p_{\rm IR}=p_\Omega = \sqrt{\frac{2 \mu v}{a}}\,.$$
Note that the IR cutoff has not been assumed, rather it was a derived output of the calculation. Together with the (again, derived) UV scale, this gives the argument of universal logarithm for SFDM media,
\begin{equation}
    \frac{p_{\rm{UV}}}  {p_{\rm{IR}}} = \sqrt{2 \mu v a} = \sqrt{\zeta}\,.
\end{equation}

The force on a body following a generic bound orbit in the $\zeta\gg1$ regime is therefore
\begin{equation}
F^{i}(t)
=\frac{4\pi G^2m^2\rho_0}{v}
\left[-\frac{v^{i}}{v}\log\sqrt{\zeta}+\mathcal O(\zeta^{0})\right]\,,
\label{eq:large-lambda-log}
\end{equation}
where the order-unity terms can be both dissipative and conservative, nonlocal in time (sensitive to the history), and nonlocal in space (sensitive to both the primary and secondary objects). Only the logarithm is insensitive to the global properties of the binary orbit.

Let us emphasize what this universal result does and does not establish. In phenomenological applications, dynamical friction on a nontrivial orbit is often modeled by evaluating a local Coulomb logarithm with appropriately chosen ultraviolet and infrared cutoffs. This is sufficient for steady rectilinear motion in a homogeneous medium. For a bound orbit, however, the logarithm captures only the local part of the force; there are generically additional $\mathcal O(1)$ contributions that depend on the global trajectory. The local approximation is therefore parametrically controlled only when the momentum interval $p_\Omega\ll p\ll2\mu v$ is broad, or equivalently when $\log\sqrt{\zeta}\gg1$. Otherwise, the nonlogarithmic contributions cannot be neglected.

A few comments are in order to compare our results with previous literature e.g.~\cite{Traykova:2023qyv,vicente2022dynamical}. In the previous work the argument of the Coulomb logarithm is given by a quantity $\Lambda$ which is defined as the ratio of the UV and IR cutoffs, $\Lambda = \frac{b_{\rm max}}{\lambda_{dB}}$, with $b_{\rm max}$ being the maximum impact parameter. For straight line motion, as considered in~\cite{Traykova:2023qyv,vicente2022dynamical} and most other work, $b_{\rm max}$ is an IR cutoff \emph{choice} that is imposed by hand to render the calculation finite. 

In contrast, for dynamical friction in particulate matter (e.g. CDM) numerical computations have been fit to find a large logarithm with an effective UV length scale $R_{90}$ (as expected) and an IR length scale $2R$~\cite{kim2007dynamical,Desjacques_2022}. In the absence of analytical results for circular motion in SFDM, it has often been assumed in the literature that $2R$ would also serve as the heuristic IR scale in this case, obtaining a Coulomb logarithm with argument $ \Lambda_{\rm Heuristic} = 2 \mu v R\equiv \zeta_{circ}$. This heuristic is actually the square of our derived result, where the IR cutoff is governed by $p_\Omega$, and thus leads to a factor of $2$ error in the predicted force. From our explicit computation, we instead find the argument of the universal log to be $\Lambda_{\rm EFT} = \sqrt{\zeta} = \sqrt{2 \mu v a}$.

% ==================================================================
\section{Ensemble-mean dynamical friction in a stochastic scalar background}
\label{sec:stochastic-mean}

We now allow the scalar background to be a general spatially and temporally varying solution of the free Schr\"{o}dinger equation. Thus, $\psi_0(t,\bm x)$ is a superposition of freely propagating waves. The problem is much richer, as the variations of the background introduce spatial and temporal coherence scales in addition to those already present in the orbit.

For rectilinear motion, the ensemble-mean drag in a random-wave background was derived in Ref.~\cite{Lancaster:2019mde}, while Ref.~\cite{Buehler:2022tmr} compared the coherent circular-orbit result with numerical calculations in a random-wave background. Here we extend the analytic treatment to a general periodic binary and determine how the mean response changes across the hierarchy of orbital and coherence scales.

In this section we derive the ensemble-mean orbit-averaged energy flux; in \cref{sec:stochastic-autocorrelator}, we study its fluctuations through a two-time correlation function.

We restrict attention to the orbit-averaged energy flux, $\overline{\FE}$, rather than the instantaneous force vector. This is both simpler to compute and the leading quantity relevant to the adiabatic evolution of a binary inspiral~\cite{Hinderer:2008dm}. Two distinct averages appear: we first time-average over one orbit in a fixed realization and then ensemble-average the resulting flux. We reserve ``mean'' and angled brackets for the ensemble average.

\subsection{Random-wave ensemble and mean response}
\label{subsec:averaged-susceptibility}

The force and flux for a fixed realization are given by the general expressions in Eqs.~\eqref{eq:chi-general}, \eqref{eq:forceinstant2general}, and \eqref{eq:orbitaveragedenergeneral}. These expressions simplify considerably after ensemble averaging. We take the ensemble to be stationary, homogeneous, and isotropic, while leaving its momentum distribution otherwise general.

A realization can be written as
\begin{equation}
\psi_0(t,\bm x)
=\int_{\bm k} a(\bm k)
 e^{-\ii E_{\bm k}t+\ii\bm k\cdot\bm x},
\qquad
E_{\bm k}=\frac{k^2}{2\mu},
\label{eq:random-wave-expansion}
\end{equation}
with two-point function
\begin{equation}
\avg{a(\bm k)a^*(\bm k')}
=(2\pi)^3\delta^3(\bm k-\bm k')P_\psi(k).
\label{eq:mode-covariance}
\end{equation}
The force is quadratic in the background field, so Eq.~\eqref{eq:mode-covariance} is sufficient for the ensemble mean; Gaussianity will only become necessary when we compute higher moments in Sec.~\ref{sec:stochastic-autocorrelator}.

It will be useful to trade the mode power spectrum for a velocity-space mass distribution.  Defining
\begin{equation}
\bm u\equiv\frac{\bm k}{\mu},
\qquad
f(\bm u)\equiv\frac{\mu^4}{(2\pi)^3}P_\psi(\mu u),
\qquad
\bar\rho=\int\dd^3u\,f(\bm u),
\end{equation}
we can discuss the random-wave response using the same velocity variables that arise naturally in kinetic theory. We do not assume a particular form for $f(\bm u)$ in this section. In \cref{sec:stochastic-autocorrelator}, we use a Maxwell--Boltzmann distribution as an explicit example.

For a fixed realization, the density response function in Eq.~\eqref{eq:chi-general} is momentum nondiagonal. Statistical translation invariance restores momentum conservation after averaging, so that $p=q=(\omega,\bm p)$ on the support of the delta function,
\begin{equation}
\avg{\chi(p,q)}
=(2\pi)^4\delta^4(p-q)\,
\bar\chi^R(\omega,\bm p).
\label{eq:chi-diagonal-average}
\end{equation}
After using $p=q$, and shifting integration variables, the density response function can be written in a particularly useful form,
\begin{equation}
\bar\chi^R(\omega,\bm p)
=
 \bm p^2\int\dd^3u\,
\frac{f(\bm u)}
{\left(\omega-\bm p\cdot\bm u+\ii0\right)^2-E_{\bm p}^2}.
\label{eq:averaged-susceptibility}
\end{equation}
Thus each occupied background mode has the same density response as a stationary coherent mode, but with the frequency Doppler shifted by $\bm p\cdot\bm u$. The ensemble mean then averages these responses over the velocity distribution.

Inserting Eq.~\eqref{eq:averaged-susceptibility} into the master flux formula, Eq.~\eqref{eq:orbitaveragedenergeneral}, gives
\begin{align}
\avg{\overline{\FE}}
=&\frac{\ii(4\pi G)^2}{T}\int_p \frac{\omega}{\bm p^2} \nonumber \\
&\times\int_{T_s-T/2}^{T_s+T/2}\dd t
\int\dd t'
\rho_{\rm bin}(-p;t)\rho_{\rm bin}(p;t') \nonumber \\
&\times
\int\dd^3u\,
\frac{f(\bm u)}
{\left(\omega-\bm p\cdot\bm u+\ii0\right)^2-E_{\bm p}^2}\,,
\end{align}
where $T_s$ is the center of the orbital averaging window. The second line isolates the source and the third isolates the environmental response. Stationarity makes the ensemble mean independent of $T_s$. For a periodic orbit, the harmonic source defined in Appendix~\ref{app:feynmandiagrams} reduces the time and frequency integrals to
\begin{align}\label{eq:mean-flux-generalcase2}
\avg{\overline{\FE}}
={}&\ii(4\pi G)^2
\sum_{n\neq0}
\int_{\bm p}
\frac{n\Omega}{p^2}
\left|\mathcal S_n(\bm p)\right|^2
\\
&\times
\int\dd^3u\,
\frac{f(\bm u)}
{\left(n\Omega-\bm p\cdot\bm u+\ii0\right)^2-E_{\bm p}^2}.
\end{align}
This form is particularly well suited to a long-wavelength multipole expansion, as in \cref{sec:small-lambda}. In the opposite regime, as in \cref{sec:large-Lambda}, high harmonics contribute and the flux can be dominated by the local tangent to the orbit, 
\begin{align}\label{eq:local-tangent-source}
    \sum_{n=-\infty}^{\infty}\left|\mathcal S_n(\bm p)\right|^2& \delta(\omega-n\Omega)\bigg|_{\rm large \, orbits} \approx\nonumber \\
    &\sum_{A}m_{A}^2\frac{1}{T}\int_{T_s-T/2}^{T_s+T/2}\dd t\,\delta(\omega-\bm p\cdot \bm v_{A}(t))\,.
\end{align}
The sum runs over the bodies which have ``large'' orbits relative to the relevant environmental wavelength.

The retarded poles in the density response function make explicit which values of $\bm p$ can contribute to dissipation.  Using
\begin{equation}
\operatorname{Im}
\frac{1}{(x+\ii0)^2-E_{\bm p}^2}
=-\frac{\pi}{2E_{\bm p}}
\left[\delta(x-E_{\bm p})-\delta(x+E_{\bm p})\right],
\end{equation}
and doing a simple relabeling, we obtain
\begin{align}\label{eq:mean-flux-generalcase}
\avg{\overline{\FE}}
={}&\pi(4\pi G)^2
\sum_{n\neq0}\int_{\bm p}
\frac{n\Omega}{p^2E_{\bm p}}
\left|\mathcal S_n(\bm p)\right|^2 \nonumber
\\
&\times
\int\dd^3u\,f(\bm u)\delta\!\left(n\Omega-\bm p\cdot\bm u-E_{\bm p}\right)\,.
\end{align}
The dissipative support therefore satisfies
\begin{equation}\label{eq:on-shell-condition}
 n\Omega=\bm p\cdot\bm u + E_{\bm p}.
\end{equation}
For $f(\bm u)=\bar\rho\,\delta^3(\bm u)$ \cref{eq:mean-flux-generalcase} reduces immediately to the coherent flux of Sec.~\ref{sec:small-lambda}.

\subsection{Independent medium and source expansions}
\label{subsec:bounded-scales}

The preceding constant-background  analysis contained two physical regimes, distinguished by the relative sizes of $\lambda_{\rm dB}$ and $R$. The velocity distribution $f(\bm u)$ introduces a third independent length scale. For generic parameters the flux is difficult to evaluate analytically, but parametrically separated scales define controlled asymptotic regimes. We characterize all eight such hierarchies below.\footnote{Three independent scales naively give six orderings. However, the geometric mean $\sqrt{\lambda_{\rm dB}R}$ splits two of them into subcases, giving eight hierarchies in total. Up to an order-unity factor, its inverse is the scale $p_\Omega$ introduced in \cref{sec:large-Lambda}.}

It is useful to identify two independent crossover scales in the momentum integral. We characterize the velocity distribution by a typical speed $u_\star$ and define
\begin{equation}
 k_\star\equiv\mu u_\star,
\qquad
\lambda_\star\equiv k_\star^{-1}=\frac{1}{\mu u_\star},
\qquad
\frac{u_\star}{v}=\frac{\lambda_{\rm dB}}{\lambda_\star}.
\end{equation}
The scale $\lambda_\star$ is the characteristic wavelength of an occupied background mode. 

The first crossover is intrinsic to the medium response.  At a fixed transfer momentum $p$, the two frequency scales in Eq.~\eqref{eq:averaged-susceptibility} are the \textit{Doppler} and \textit{recoil} frequencies,
\begin{equation}
 \omega_{\rm D}(p)\sim p u_\star,
\qquad
 \omega_{\rm rec}(p)=E_{\bm p}=\frac{p^2}{2\mu},
\qquad
\frac{\omega_{\rm rec}}{\omega_{\rm D}}
\sim\frac{p}{2k_\star}.
\end{equation}
For $p\ll k_\star$ the medium responds kinetically (i.e., like a uniform field of velocity-distributed free particles), and for  $k_\star\ll p$  the medium responds coherently. In the kinetic region recoil can be neglected and
\begin{align}\label{eq:kin-resp-approx}
\bar\chi^R_{\rm kin}(\omega,\bm p)
&\simeq
p^2\int\dd^3u\,
\frac{f(\bm u)}{(\omega-\bm p\cdot\bm u+\ii0)^2}
\notag\\
&=-\int\dd^3u\,
\frac{\bm p\cdot\bm\nabla_{u}f(\bm u)}
{\omega-\bm p\cdot\bm u+\ii0},
\end{align}
where the second form follows by integration by parts.  The full velocity distribution remains resolved.  In the opposite limit, $k_\star\ll p$, the Doppler shift is subleading and
\begin{equation}\label{eq:coh-resp-approx}
\bar\chi^R_{\rm coh}(\omega,\bm p)
\simeq
\frac{\bar\rho\,p^2}
{(\omega+\ii0)^2-E_{\bm p}^2}\,.
\end{equation}
At leading order the detailed distribution has disappeared and only the mean density remains.

The second crossover is set by the source.  The binary factors contain phases $e^{-\ii\bm p\cdot\bm x_A}$, so
\begin{equation}
\begin{aligned}
 pR\ll1
 &\quad\Longrightarrow\quad
 \text{multipolar source},
\\
 pR\gg1
 &\quad\Longrightarrow\quad
 \text{locally resolved source}.
\end{aligned}
\end{equation}
Thus the integrand contains four possible asymptotic sectors,
\begin{equation}
\begin{array}{c|cc}
 & pR\ll1 & pR\gg1 \\
\hline
p\ll k_\star & \text{multipolar kinetic} & \text{local kinetic} \\
p\gg k_\star & \text{multipolar coherent} & \text{local coherent}
\end{array}.
\end{equation}
These labels characterize regions of the momentum integral. Since the flux requires an integral over $p$, to determine which sector controls a given hierarchy we must ask where the dissipative support actually lies and solve the on-shell condition,~\cref{eq:on-shell-condition}.

For systems well described by the coherent regime, the response function in Eq.~\eqref{eq:coh-resp-approx} is precisely the one used in the preceding sections.\footnote{The fixed background density $\rho_0$ is replaced by the ensemble-mean density $\bar\rho$.} The results of \cref{sec:small-lambda,sec:large-Lambda} therefore cover these cases at leading order in the relevant dimensionless ratios.

We do not explicitly evaluate the flux integrals in the kinetic cases. Dynamical friction in kinetic systems is well studied, dating back to Chandrasekhar's calculation, while our focus is the distinctive behavior of matter-wave environments. The kinetic calculations nevertheless follow directly from the present setup: one inserts the kinetic response in Eq.~\eqref{eq:kin-resp-approx} into the mean-flux formula, Eq.~\eqref{eq:mean-flux-generalcase2}. In the multipolar regime, the source $\mathcal S_n(\bm p)$ can be expanded as in \cref{subsec:binary-multipoles,subsec:eccentric-quadrupole}; in the local regime, the harmonic sum can be replaced by its large-$n$ local-tangent approximation, Eq.~\eqref{eq:local-tangent-source}.

In the following subsections we will characterize the system according to its dimensionless ratios, and we will map out for which values of these ratios the appropriate description is kinetic or coherent and multipolar or local. 

\subsection{Dissipative support}

We now address the following question: for a given system, what values of the momentum transfer satisfy the on-shell condition? The answer is qualitatively different between the multipolar and locally resolved sources.  We therefore treat the two source limits separately before introducing the complete hierarchy map.

\subsubsection{Multipolar source: low orbital harmonics}\label{sec:multipolar-low-n}

For $pR\ll1$, the full binary source can be multipole expanded. As shown in \cref{subsec:binary-multipoles}, the leading time-dependent source is quadrupolar,
\begin{equation}
\mathcal S_n(\bm p)
=-\frac{1}{2}p_i p_j I_n^{ij}
+\mathcal O\!\left((pR)^3\right),
\qquad n\neq0,
\end{equation}
and the harmonics which dominate the sum are $n=\mathcal O(1)$. As a result, the on-shell condition can be analyzed directly.

If the on-shell condition is dominated by the Doppler term, then the momentum $p$ is of the order
\begin{equation}
 p_D\sim\frac{n\Omega}{u_\star}.
\end{equation}
If instead it is dominated by scalar recoil, the momentum is
\begin{equation}
 p_{\rm rec}\sim\sqrt{2\mu n\Omega}.
\end{equation}
It is convenient to introduce the bounded-orbit scale
\begin{equation}
 L_\Omega\equiv(\mu\Omega)^{-1/2}
 \sim\sqrt{R\lambda_{\rm dB}}.
\label{eq:Lomega}
\end{equation}
For $n=\mathcal O(1)$, the Doppler-dominated kinetic solution obeys
\begin{equation}
\frac{p_D}{k_\star}
\sim n\frac{\lambda_\star^2}{R\lambda_{\rm dB}}
\sim n\left(\frac{\lambda_\star}{L_\Omega}\right)^2,
\qquad
p_D R\sim n\frac{\lambda_\star}{\lambda_{\rm dB}},
\end{equation}
whereas the recoil-dominated coherent solution satisfies
\begin{equation}
\frac{k_\star}{p_{\rm rec}}
\sim\frac{1}{\sqrt{2n}}\frac{L_\Omega}{\lambda_\star},
\qquad
p_{\rm rec}R\sim\sqrt{\frac{2nR}{\lambda_{\rm dB}}}.
\end{equation}
The dimensionless ratios given for each regime must be parametrically small in order for the system to be well-described by that regime.

The above relations make the regime classification immediate, in this $pR\ll1$ multipolar limit.  A Doppler-dominated low-harmonic contribution can be simultaneously kinetic and multipolar when
\begin{equation}\label{eq:kinetic-multipolar-conditions}
 \lambda_\star\ll L_\Omega,
\qquad
 \lambda_\star\ll\lambda_{\rm dB},
\end{equation}
while a recoil-dominated contribution is simultaneously coherent and multipolar when
\begin{equation}
 L_\Omega\ll\lambda_\star,
\qquad
 R\ll\lambda_{\rm dB}\,,
\end{equation}
in which case the mean response reduces directly to the coherent calculation of Sec.~\ref{sec:small-lambda} evaluated at $\bar\rho$.

The first set of conditions, \cref{eq:kinetic-multipolar-conditions}, require one qualification.  The fact that the Doppler-dominated condition $p_D\ll k_\star$ is satisfied establishes that there exists support within the $p$ integral for which the on-shell condition is satisfied and the response is kinetic ($p\ll k_{\star}$). It does not, however, prove that the full integral is localized for $p\sim p_D$. Depending on the shape of the orbit it may be the case that the integral also receives contributions from the crossover region $p\sim k_{\star}$, in which case the response is not definitively kinetic or coherent, and the complete density response function needs to be retained for computations. This subtlety does not affect the other regimes we discuss, nor any explicit computations that we perform.

\subsubsection{Locally resolved source: a broad momentum region}

When $pR\gg1$, the low-harmonic reasoning above is no longer appropriate.  As in Sec.~\ref{sec:large-Lambda}, the rapidly varying source is instead controlled by short time separations over which each worldline is locally straight.  Equivalently, the large-$pR$ harmonic sum contains harmonics with $n\sim pR$, so that locally
\begin{equation}
 \omega\simeq\bm p\cdot\bm v.
\end{equation}
The on-shell condition controlling the support of the flux integral then becomes
\begin{equation}
 \bm p\cdot(\bm v-\bm u)
 = E_{\bm p}\,.
\end{equation}
For a background component with relative velocity
\begin{equation}
 \bm w\equiv\bm v-\bm u,
\end{equation}
the allowed momentum transfer extends over a broad interval up to $p\sim2\mu w$.  The local problem therefore does not select a unique analogue of $p_D$ or $p_{\rm rec}$; it samples a range of momentum transfers, precisely as in the hard-region analysis of Sec.~\ref{sec:large-Lambda}.

There is nevertheless a simple physical distinction.  If the bath is relatively ``fast/hot'', $u_\star\gg v$, or equivalently $\lambda_\star\ll\lambda_{\rm dB}$, the relative motion is controlled by the bath and the mean local response must retain the velocity distribution.  The leading local drag is then kinetic. 

If the bath is relatively ``slow/cold'', $u_\star\ll v$, or $\lambda_{\rm dB}\ll\lambda_\star$, at leading order we can neglect $\bm p \cdot \bm u$ and the mean flux reduces to the coherent local response evaluated at $\bar\rho$.  

Schematically, the local mean can be viewed as the velocity average
\begin{equation}
 \avg{\overline{\FE}}_{\rm local}
 \simeq
 \int\dd^3u\,\frac{f(\bm u)}{\bar\rho}\,
 \mathcal F_{E,\rm coh}^{\rm line}(\bar\rho,\bm v-\bm u),
\end{equation}
with the appropriate sum over the two bodies understood. Here $\mathcal F_{E,\rm coh}^{\rm line}(\rho,\bm w)$ denotes the steady energy flux for rectilinear motion with relative velocity $\bm w$ through a coherent background of density $\rho$.  This makes the two limits transparent: a fast bath requires the full convolution, while a slow bath permits an expansion around $\bm u=0$.

\subsubsection{Kinetic response and the ultraviolet cutoff}

In the local kinetic regimes the mean-force calculation has the same structure as a collisionless velocity-space calculation: the response is averaged over the environmental distribution $f(\bm u)$.  This does not mean that wave physics is completely irrelevant: at very short distances the $p^{2}/2\mu$ recoil term can eventually become relevant, and this has the effect of softening the would-be UV divergence.

For a background component with relative speed
\begin{equation}
 w=|\bm v-\bm u|,
\end{equation}
define the relative response wavelength and the classical strong-deflection scale,
\begin{equation}
 \lambda_{\rm resp}(w)\sim\frac{1}{\mu w},
\qquad
 R_{90}(w)\sim\frac{Gm}{w^2}.
\end{equation}
If $\lambda_{\rm resp}\gg R_{90}$, wave behavior regulates the weak-scattering integral before classical strong deflection is reached.  If $\lambda_{\rm resp}\ll R_{90}$, the medium is effectively particle-like down to the strong-scattering scale.  Parametrically, the lower impact scale entering a leading logarithm is therefore
\begin{equation}
 b_{\min}^{\rm eff}(w)
 \sim
 \max\!\left[\lambda_{\rm resp}(w),R_{90}(w),R_{\rm object},\ldots\right].
\end{equation}

For a broad velocity distribution this crossover must remain inside the $\bm u$ integral.  Moreover, when $b$ approaches $R_{90}$ linear response itself ceases to be controlled and a matching calculation is required.

\subsection{Length-hierarchy map}
\label{subsec:six-regimes}

We can now translate the momentum-space criteria into a hierarchy of physical lengths.  The scale $L_\Omega$ is the geometric mean of the orbital size and the de Broglie wavelength, and therefore always lies between them,
\begin{equation}
 \min(R,\lambda_{\rm dB})
 <L_\Omega<
 \max(R,\lambda_{\rm dB}).
\end{equation}
Consequently, if $\lambda_\star$ lies outside the interval bounded by $R$ and $\lambda_{\rm dB}$, its ordering relative to $L_\Omega$ is automatic.  When it lies between these scales though, the ordering of $\lambda_\star$ and $L_\Omega$ distinguishes sub-cases.

There are six possible orderings of the three basic lengths $(R,\lambda_{\rm dB},\lambda_\star)$, and the orderings $R<\lambda_\star<\lambda_{\rm dB}$ and $\lambda_{\rm dB}<\lambda_\star<R$ split at $L_\Omega$, giving eight physically distinct hierarchies to map out. 

It is most transparent to organize them first by the two hierarchies familiar from the coherent case, $R<\lambda_{\rm dB}$ and $\lambda_{\rm dB}<R$. We then scan $\lambda_\star$ monotonically from short to long wavelength within each branch. Table~\ref{tab:six-regimes} summarizes the results.

Before the case-by-case discussion, we introduce one more property beyond the source and response classifications. A realization of the random background varies in both space and time, so the orbit may sample many distinct background configurations. The amount of spatial and temporal sampling is an independent characteristic of each regime.

To describe this sampling question, consider a broadly occupied spectrum with momentum width $\Delta k\sim k_\star$. Its intrinsic dephasing time is
\begin{equation}
 t_{\rm coh}\sim\frac{1}{\mu u_\star^2}
 =\mu\lambda_\star^2,
 \qquad
 \frac{t_{\rm coh}}{T}
 \sim\frac{\lambda_\star^2}{R\lambda_{\rm dB}}
 \sim\left(\frac{\lambda_\star}{L_\Omega}\right)^2.
\label{eq:coherence-time}
\end{equation}
This estimate depends on the width of the distribution $f(\bm u)$; a narrow shell has a parametrically longer coherence time. The time it takes an orbiting body to cross one spatial interference patch instead scales as
\begin{equation}
 t_{\rm cross}\sim\frac{\lambda_\star}{v},
 \qquad
 \frac{t_{\rm cross}}{T}\sim\frac{\lambda_\star}{R}.
\label{eq:crossing-time}
\end{equation}
These relations determine the sampling column of Table~\ref{tab:six-regimes}. They provide useful intuition for the fluctuation problem, but do not by themselves establish self-averaging; we return to that question in Sec.~\ref{sec:stochastic-autocorrelator}.

\begin{table*}[t]
\caption{Regimes for the ensemble-mean orbit-averaged energy flux. $\mathcal O(1)$ numerical factors are suppressed. The six orderings of $(R,\lambda_{\rm dB},\lambda_\star)$ produce eight entries because the two intermediate-$\lambda_\star$ hierarchies, $R<\lambda_\star<\lambda_{\rm dB}$ and $\lambda_{\rm dB}<\lambda_\star<R$, both split at $L_\Omega$. Since $L_\Omega$ always lies between $R$ and $\lambda_{\rm dB}$, we do not write it explicitly unless needed to distinguish subcases. The sampling column assumes a broad spectrum of background modes, $\Delta k\sim k_\star$, and describes the background textures (densities and velocities) sampled in one orbit.}
\label{tab:six-regimes}
\footnotesize
\begin{tabular}{@{}p{0.045\textwidth} p{0.235\textwidth} p{0.20\textwidth} p{0.20\textwidth} p{0.25\textwidth}@{}}
\toprule
No. & Hierarchy & Mean medium response & Source treatment & Sampling over an orbit in one realization \\
\midrule
1 & $\lambda_\star\ll R\ll\lambda_{\rm dB}$
& kinetic; retain $f(\bm u)$
& local/rectilinear
& many spatial patches; rapidly evolving \\
2 & $R\ll\lambda_\star\ll L_\Omega\ll\lambda_{\rm dB}$
& partly kinetic; see caveat
& multipole
& one spatial patch, \textbf{but} rapidly evolving \\
3 & $R\ll L_\Omega\ll\lambda_\star\ll\lambda_{\rm dB}$
& coherent
& multipole
& one patch; quasi-static \\
4 & $R\ll\lambda_{\rm dB}\ll\lambda_\star$
& coherent
& multipole
& one patch; quasi-static \\
5 & $\lambda_\star\ll\lambda_{\rm dB}\ll R$
& kinetic; retain $f(\bm u)$
& local/rectilinear
& many spatial patches; rapidly evolving \\
6 & $\lambda_{\rm dB}\ll\lambda_\star\ll L_{\Omega}\ll R$
& coherent, at leading order
& local/rectilinear
& many spatial patches; rapidly evolving\\
7 & $\lambda_{\rm dB}\ll L_{\Omega}\ll\lambda_\star\ll R$
& coherent, at leading order
& local/rectilinear
& many spatial patches, \textbf{but} quasi-static \\
8 & $\lambda_{\rm dB}\ll R\ll\lambda_\star$
& coherent, at leading order
& local/rectilinear
& one patch; quasi-static \\
\bottomrule
\end{tabular}
\end{table*}

\subsubsection*{Case 1: $\lambda_\star\ll R\ll\lambda_{\rm dB}$}

The bath is fast, $(\lambda_\star\ll \lambda_{\rm dB})$, and its characteristic wavelength is shorter than the orbit $(\lambda_\star\ll R)$. It thus responds quickly to the local straight-line source, before resolving the curvature of the orbit. A broad momentum interval lies in the kinetic regime $(p^{-1}\gg \lambda_{\star})$, and the leading mean drag must retain the velocity distribution.  The nominal orbital de Broglie wavelength is long, but it does not set the response scale because the relative motion is controlled by the bath. This case is then local-kinetic. 

\subsubsection*{Case 2: $R\ll\lambda_\star\ll L_\Omega\ll\lambda_{\rm dB}$}

The orbit lies inside one background patch $(R\ll\lambda_{\star})$, so the source is multipolar. The bath is nevertheless fast $(\lambda_{\star}\ll\lambda_{\rm dB})$, and the low orbital harmonics admit a Doppler-dominated solution with
\begin{equation}
 p_D R\sim\frac{\lambda_\star}{\lambda_{\rm dB}}\ll1,
\qquad
\frac{p_D}{k_\star}
\sim\left(\frac{\lambda_\star}{L_\Omega}\right)^2\ll1.
\end{equation}
Thus, a controlled kinetic/Doppler part of the dissipative support exists. However, this is the subtle case discussed at the end of \cref{sec:multipolar-low-n}. These inequalities do not establish that the complete quadrupole-weighted momentum integral is dominated by $p\sim p_D$. Since the source remains multipolar through $p\sim k_\star$, we retain Eq.~\eqref{eq:averaged-susceptibility} without expanding the medium response across this crossover. This case is therefore partly multipolar-kinetic and may also receive multipolar contributions for which neither the kinetic nor coherent approximation is controlled.

\subsubsection*{Case 3: $R\ll L_\Omega\ll\lambda_\star\ll\lambda_{\rm dB}$}

The source is again multipolar ($R\ll\lambda_{\star}$), but the low-harmonic on-shell momentum now lies on the coherent side of the medium crossover,
\begin{equation}
 p_{\rm rec}R\sim\sqrt{\frac{R}{\lambda_{\rm dB}}}\ll1,
\qquad
\frac{k_\star}{p_{\rm rec}}\sim\frac{L_\Omega}{\lambda_\star}\ll1\,.
\end{equation}
 The velocity distribution is therefore unresolved and the ensemble mean reduces to the coherent-multipolar case considered in Sec.~\ref{sec:small-lambda}, with $\rho_0\rightarrow\bar\rho$. 

\subsubsection*{Case 4: $R\ll\lambda_{\rm dB}\ll\lambda_\star$}

This is the clean slow-bath coherent-multipolar limit. Because $L_\Omega$ lies between $R$ and $\lambda_{\rm dB}$, the hierarchy automatically implies $L_\Omega\ll\lambda_\star$. The recoil momentum satisfies $p_{\rm rec}R\ll1$ and $k_\star\ll p_{\rm rec}$, so both expansions are parametrically controlled. This case again reduces to the coherent-multipolar result evaluated at $\bar\rho$.

\subsubsection*{Case 5: $\lambda_\star\ll\lambda_{\rm dB}\ll R$}

The bath is fast and the source is locally resolved.  Here $u_\star\gg v$, so the relative velocity is set by the bath and the local response retains the full distribution $f(\bm u)$.  The leading hard contribution is the velocity-distributed analogue of the local straight-line response discussed in Sec.~\ref{sec:large-Lambda}. This case is local-kinetic. 

\subsubsection*{Case 6: $\lambda_{\rm dB}\ll\lambda_\star\ll L_{\Omega}\ll R$}

The source is still locally resolved, but now the orbital motion is faster than the bath, $(v\gg u_\star)$.  At leading order in $u_\star/v$ we can omit the $\bm p\cdot \bm u$ Doppler contribution. The leading ensemble mean therefore reduces to the local-coherent response at $\bar\rho$. 

\subsubsection*{Case 7: $\lambda_{\rm dB}\ll L_{\Omega}\ll\lambda_\star\ll R$}

The source is locally resolved, and the bath is relatively slow, so this is the same local-coherent response regime as Case 6.

\subsubsection*{Case 8: $\lambda_{\rm dB}\ll R\ll\lambda_\star$}

This is the clean slow-bath local-coherent limit, effectively equivalent to the computation in~\cref{sec:large-Lambda}. The bath is slow, as in Case 6, so its velocity dispersion is a perturbation and the leading local mean is the coherent response at $\bar\rho$. The source remains local, even though the orbit lies inside a single background patch, because the relevant short response scale is now the orbital de Broglie wavelength, $\lambda_{\rm dB}\ll R$, rather than $\lambda_\star$. The bath velocity distribution is therefore irrelevant to the mean force at leading order. This case is local-coherent.

Now that we've mapped out the eight different regimes and provided explicit expressions for many of the cases, we'd like to have a more refined understanding of the statistical properties of the force. In the following section we study the fluctuations about this mean value, both across the ensemble and as a function of time for a single realization.

% ==================================================================
\section{Realization-dependent dynamical friction: fluctuations and correlations}
\label{sec:stochastic-autocorrelator}

In the previous section we characterized the ensemble-mean orbit-averaged energy flux from a binary in a stochastic scalar background.  We now ask a distinct question: how well does that ensemble mean represent the flux in one realization? The appropriate object to compute is the connected two-time correlator of the orbit-averaged flux.  At equal times it gives the realization-to-realization variance, while at separated times it determines the correlation time of the stochastic dissipation from the binary.

The coherent/kinetic classification of the \emph{mean response} does not, by itself, answer this question. One might expect that the coherent cases see the background as adiabatically evolving, while the kinetic cases see the background as rapidly self-averaging. This intuition is valuable, but as we will demonstrate, not quite correct. 

The rate at which the orbiting bodies sample different background textures is controlled by the beating of off-diagonal phases between occupied background modes. Cases 3, 4, and 8 are coherent and sample one quasi-static patch. Cases 6 and 7 are also coherent at leading order, but the orbit crosses many patches and their leading local contribution self-averages. Conversely, rapid phase variation in the kinetic cases 1, 2, and 5 can suppress the fluctuations, but does not by itself guarantee that they are small relative to the mean flux. We therefore keep the medium response and sampling questions separate.

\subsection{Background-density correlations}
\label{subsec:background-density-correlator}

It is useful to first describe the fluctuations intrinsic to the background before it is perturbed by the binary. Using the random-wave expansion in Eq.~\eqref{eq:random-wave-expansion}, the density fluctuation $\delta\rho(t,\bm x)=\rho(t,\bm x)-\bar\rho$ is
\begin{align}
\delta\rho(t,\bm x)
={}&\mu\int_{\bm k,\bm k'}
\bigg[
a(\bm k)a^*(\bm k')
-\avg{a(\bm k)a^*(\bm k')}
\bigg]
\nonumber\\
&\times
e^{-\ii(E_{\bm k}-E_{\bm k'})t
  +\ii(\bm k-\bm k')\cdot\bm x}.
\label{eq:density-fluctuation-modes}
\end{align}
This quantity is quadratic in the background modes, as is the energy flux considered below. Assuming Gaussian statistics, Wick's theorem gives
\begin{align}
C_\rho(\tau,\bm r)
&\equiv
\avg{\delta\rho(t,\bm x)
\delta\rho(t',\bm x')}
\nonumber\\
&=\mu^2\left|
\int_{\bm k}P_\psi(k)
e^{-\ii E_{\bm k}\tau+\ii\bm k\cdot\bm r}
\right|^2,
\label{eq:density-correlation-general}
\end{align}
where $\tau=t-t'$ and $\bm r=\bm x-\bm x'$.  The disconnected contraction has canceled against the subtracted mean in Eq.~\eqref{eq:density-fluctuation-modes}.

For an explicit example, consider the isotropic Maxwell--Boltzmann distribution
\begin{equation}
f(\bm u)
=\frac{\bar\rho}{(2\pi u_\star^2)^{3/2}}
e^{-u^2/(2u_\star^2)},
\label{eq:maxwellian-background}
\end{equation}
where $u_\star$ is the one-dimensional velocity dispersion. For this distribution, define
\begin{equation}
t_{\rm coh}\equiv\frac{1}{\mu u_\star^2}.
\end{equation}
Together with $\lambda_\star=(\mu u_\star)^{-1}$ defined above, Eq.~\eqref{eq:density-correlation-general} becomes
\begin{equation}
\frac{C_\rho(\tau,\bm r)}{\bar\rho^2}
=\frac{1}{[1+(\tau/t_{\rm coh})^2]^{3/2}}
\exp\!\left[
-\frac{r^2/\lambda_\star^2}
{1+(\tau/t_{\rm coh})^2}
\right].
\label{eq:maxwellian-density-correlation}
\end{equation}
The Maxwellian correlator now makes explicit the spatial and temporal sampling estimates used in~\cref{tab:six-regimes}. There is an essential takeaway from this expression; at coincident points
\begin{equation}
\avg{\delta\rho(t,\bm x)^2}=\bar\rho^2.
\label{eq:order-one-density-fluctuations}
\end{equation}
The relative \emph{density fluctuations are order unity} even though the scalar field itself is Gaussian.  

Since the background density fluctuations are intrinsically order unity the orbit-averaged flux may inherent large fluctuations.  These flux fluctuations can, however, be dynamically suppressed by the spacetime region sampled by the orbit or by the medium's response to the perturbation.

Equation~\eqref{eq:maxwellian-density-correlation} provides useful general intuition, even though its explicit form was derived for a Maxwellian distribution.  For a broad distribution satisfying
\begin{equation}
\avg{|\bm u|}
\sim
\avg{|\bm u - \bm u'|}
\sim u_\star,
\end{equation}
the stochastic background varies appreciably over spatial scales $\lambda_\star$ and temporal scales $t_{\rm coh}$.  A binary with $R/\lambda_\star\gg1$ passes through many distinct spatial patches over one orbit, while a binary with
\begin{equation}
\frac{T}{t_{\rm coh}}
\sim
\left(\frac{L_\Omega}{\lambda_\star}\right)^2
\gg1
\end{equation}
sees the background evolve significantly during one orbital period.  The sampling properties of each of the eight hierarchies are summarized in \cref{tab:six-regimes}.

\subsection{Energy-flux autocorrelator}
\label{subsec:energy-flux-autocorrelator}

For a fixed realization, the local background evolves in time and the flux need not be stationary along a prescribed orbit. We introduce a center time $T_s$ and define the orbit-averaged flux
\begin{equation}
\overline{\FE}(T_s)
\equiv
-\frac{1}{T}\sum_{A=1}^2
\int_{T_s-T/2}^{T_s+T/2}\dd t\,
\dot{\bm x}_A(t)\cdot\bm F_A(t)\,.
\label{eq:orbit-averaged-random-flux}
\end{equation}
The overbar denotes time averaging, while angled brackets denote an ensemble average.  We define
\begin{equation}
\delta\overline{\FE}(T_s)
=\overline{\FE}(T_s)-\avg{\overline{\FE}(T_s)}\,,
\end{equation}
and the correlation function which describes these fluctuations is
\begin{equation}
C_{\FE}(T_s,T_s')
\equiv
\avg{\delta\overline{\FE}(T_s)
\delta\overline{\FE}(T_s')}.
\label{eq:flux-autocorrelator}
\end{equation}

Before ensemble averaging, the force is quadratic in the background field.  We may therefore write
\begin{equation}
\overline{\FE}(T_s)
=\int_{\bm k,\bm k'}
 a(\bm k)a^*(\bm k')
\mathcal K_{T_s}(\bm k,\bm k'),
\label{eq:flux-quadratic-kernel}
\end{equation}
where the energy-flux kernel, $\mathcal K_{T_s}$, contains the two-body source, the retarded scalar response, and the orbital average.  Reality implies
\begin{equation}
\mathcal K_{T_s}(\bm k,\bm k')^*
=\mathcal K_{T_s}(\bm k',\bm k).
\end{equation}

The mean flux is an integral over the diagonal elements,
\begin{equation}
\avg{\overline{\FE}(T_s)}
=\intk P_\psi(k)
\mathcal K_{T_s}(\bm k,\bm k)\,.
\label{eq:kernel-mean-flux}
\end{equation}
A constant background of density $\rho_0$ corresponds to $P_\psi(k)=(\rho_0/\mu)(2\pi)^3\delta^3(\bm k)$, and hence
\begin{equation}
\overline{\FE}_{\rm coh}(\rho_0)
=\frac{\rho_0}{\mu}
\mathcal K_{T_s}(\bm 0,\bm 0)\,.
\label{eq:kernel-coherent-flux}
\end{equation}
In a coherent-response regime the momenta internal to the response kernel are hard compared with $k_\star$, so its diagonal may be expanded about $\bm k=\bm 0$.  Equation~\eqref{eq:kernel-mean-flux} then reduces to the coherent result at $\bar\rho$. In contrast, in a kinetic-response regime the variation of the kernel across the occupied mode spectrum cannot be ignored.

For a Gaussian ensemble with covariance~\eqref{eq:mode-covariance} and no anomalous $\avg{aa}$ correlator, Wick's theorem gives
\begin{equation}
C_{\FE}(T_s,T_s')
=\int_{\bm k,\bm k'}
P_\psi(k)P_\psi(k')
\mathcal K_{T_s}(\bm k,\bm k')
\mathcal K_{T_s'}(\bm k',\bm k).
\label{eq:kernel-flux-covariance}
\end{equation}
This expression cleanly separates the random-wave statistics from the deterministic retarded energy-flux kernel.

\subsubsection{Explicit harmonic kernel}
\label{subsubsec:explicit-flux-kernel}

We now make the kernel explicit by inserting the random-wave expansion into Eqs.~\eqref{eq:chi-general} and~\eqref{eq:orbitaveragedenergeneral}. The resulting energy- and momentum-conserving delta functions can be used to eliminate the momentum $q$. Defining
\begin{equation}
\Delta E\equiv E_{\bm k}-E_{\bm k'},
\qquad
\Delta\bm k\equiv\bm k-\bm k',
\qquad
\bm u\equiv\frac{\bm k}{\mu},
\quad
\bm u'\equiv\frac{\bm k'}{\mu}.
\end{equation}
they set $q^0=p^0+\Delta E$ and $\bm q=\bm p+\Delta\bm k$. To further simplify the expression before we present it, we use the source harmonics $\mathcal S_n$ defined in Eq.~\eqref{eq:binary-harmonic-source} which simplify the time integrals. The result is
\begin{align}
&\mathcal K_{T_s}(\bm k,\bm k')
=\ii(4\pi G)^2\mu^2
\sum_{n,m=-\infty}^{\infty}
\int_{\bm p}
\frac{n\Omega+\Delta E}
{p^2|\bm p+\Delta\bm k|^2}
\nonumber\\
&\qquad\times
\mathcal S_n(\bm p)
\mathcal S_m(\bm p+\Delta\bm k)^*
\mathcal R_n(\bm p;\bm k,\bm k')
\nonumber\\
&\qquad\times
e^{-\ii\omega_{nm}T_s}W_T(\omega_{nm}),
\label{eq:explicit-harmonic-flux-kernel}
\end{align}
where the retarded response is packaged into
\begin{align}
\mathcal R_n(\bm p;\bm k,\bm k')
={}&
\frac{1}{n\Omega-\bm p\cdot\bm u-E_{\bm p}+\ii0}
\nonumber\\
&-
\frac{1}{n\Omega-\bm p\cdot\bm u'+E_{\bm p}+\ii0}\,,
\label{eq:harmonic-response-factor}
\end{align}
the beat frequency is
\begin{equation}
\omega_{nm}=\Delta E+(n-m)\Omega\,,
\end{equation} 
and the orbital averaging produces the window function
\begin{equation} 
W_T(\omega)
\equiv\frac{1}{T}\int_{-T/2}^{T/2}\dd t\,e^{-\ii\omega t}
=\frac{2\sin(\omega T/2)}{\omega T}\,.
\label{eq:finite-time-window}
\end{equation}
Equation~\eqref{eq:explicit-harmonic-flux-kernel} is valid for a general periodic orbit. 

The window function $W_T$ makes explicit that time-averaging is controlled by the combined background and orbital beat frequency $\Delta E+(n-m)\Omega$. The corresponding spatial beat phase is instead contained in
\begin{equation}
\mathcal S_m(\bm p+\Delta\bm k)^*
=
\frac{1}{T}\int_0^T\dd t\,
e^{-\ii m\Omega t}
\sum_A m_A
e^{\ii\bm p\cdot\bm x_A(t)}
e^{\ii\Delta\bm k\cdot\bm x_A(t)}.
\label{eq:shifted-source-spatial-beat}
\end{equation}
The factor $e^{\ii\Delta\bm k\cdot\bm x_A(t)}$ controls how many spatial patches are sampled.

Together, the characteristic magnitude of the beat phase is
\begin{align}
    \Phi_{\rm beat}(t) &= \Delta\bm k\cdot \bm x_A(t)-\Delta E t, \nonumber \\
    |\Phi_{\rm beat}|_{\rm typ}
    &\sim \frac{R}{\lambda_\star}+\left(\frac{L_\Omega}{\lambda_\star}\right)^2,
\end{align}
whose size controls the coherence of the background, as perceived by the binary.

\subsection{Fluctuations in coherent-response regimes}
\label{subsec:coherent-flux-fluctuations}

 \subsubsection{One quasi-static patch: cases 3, 4, and 8}

In cases 3, 4, and 8, both $R/\lambda_{\star}\ll1$ and $T/t_{\rm coh}\ll1$, so the orbit remains within one quasi-static background patch. The occupied background modes are also soft compared with the response momenta. As argued above, the ensemble-mean flux in each of these regimes therefore reduces, at leading order, to the coherent flux computed in a constant background of density $\bar\rho$. We denote the previously obtained result for the $i^{\rm th}$ case by $\overline{\FE}^{(i)}_{\rm coh}(\bar\rho)$.

At the order considered here, the coherent flux is linear in the background density. It is therefore useful to define
\begin{equation}
\gamma_i
\equiv
\frac{\overline{\FE}^{(i)}_{\rm coh}(\bar\rho)}{\bar\rho},
\end{equation}
so that $\overline{\FE}^{(i)}_{\rm coh}(\rho)=\gamma_i\rho$ for a constant background of density $\rho$. The question for an individual stochastic realization is then whether $\bar\rho$ may be replaced by the density of the particular patch containing the orbit. Before making this replacement, however, we must ensure that the patch remains effectively unchanged over the full time over which the force integral is evaluated.

In general, the force is hereditary: the force at a given time depends on the trajectory over some interval $t_{\rm mem}$ into the past. Although this memory integral localizes explicitly for the universal leading-logarithmic contribution in the local regimes, it need not do so for the full response, particularly in the multipolar cases.

We do not calculate $t_{\rm mem}$ in this work. Instead, we make the working assumption
\begin{equation}
t_{\rm mem}={\rm few}\times\mathcal O(T),
\label{eq:memory-time-assumption}
\end{equation}
so that the force becomes insensitive to the earlier motion after a few orbital periods. This assumption is well-motivated by the finite-time calculations by Buehler and Desjacques~\cite{Buehler:2022tmr}, but we have not explicitly established it in this work. Since $T\ll t_{\rm coh}$ in cases 3, 4, and 8, it implies $t_{\rm mem}\ll t_{\rm coh}$. The background density and beat phase can therefore be treated as constant over the portion of the orbital history relevant to the force.

The response is multipolar in cases 3 and 4 and local in case 8, but in all three cases the random background consequently enters at leading order only through the density of the patch containing the orbit. Thus
\begin{equation}
\overline{\FE}^{(i)}(T_s)
\simeq
\gamma_i\rho(T_s,\bm x_c),
\qquad i=3,4,8,
\label{eq:single-patch-local-density-flux}
\end{equation}
where $\bm x_c$ denotes the center of the orbit. Any point within the orbit is equivalent at this order because $R\ll\lambda_\star$. Consequently, defining $\Delta T\equiv T_s-T_s'$, we have
\begin{equation}
C_{\FE}^{(i)}(\Delta T)
\simeq
\gamma_i^2 C_\rho(\Delta T,\bm 0),
\qquad i=3,4,8.
\label{eq:local-density-covariance-hypothesis}
\end{equation}
For the Maxwellian example, the normalized two-point flux correlation function is
\begin{equation}
\frac{C_{\FE}^{(i)}(\Delta T)}
{\avg{\overline{\FE}^{(i)}}^2}
\simeq
\frac{1}{[1+(\Delta T/t_{\rm coh})^2]^{3/2}}.
\label{eq:single-patch-flux-correlation}
\end{equation}
At equal times, the rms fluctuation is therefore of the same order as the mean. Thus cases 3, 4, and 8 exhibit no orbital self-averaging at leading order.

This result has two immediate implications. First, the local density varies by an order-one amount between realizations. Knowledge of the characteristic ensemble parameters $\bar\rho$ and $u_\star$ is therefore insufficient to predict the dynamical friction force acting on an individual binary, unless we know additional details about its local environment. The ensemble-mean prediction carries an irreducible realization-dependent uncertainty of order unity.

Second, the force remains correlated over a time of order $t_{\rm coh}$. If the inspiral timescale $t_{\rm insp}$ is long,
\begin{equation}
T\ll t_{\rm coh}\ll t_{\rm insp},
\end{equation}
then an inspiral samples many statistically distinct patches over time. We emphasize, this is a physically distinct regime from the self-averaging discussed above.  Instead, in this case the binary's orbital parameters will evolve in time like a ``Brownian'' process,  with stochastic drift and diffusion.

We defer the propagation of these stochastic force fluctuations into orbital parameters and waveform phases to future work.

\subsubsection{Many spatial patches: cases 6 and 7}

Cases 6 and 7 are coherent-response regimes, but they have $R\gg\lambda_\star$, so each body moves through many background patches in a single orbit. In these cases, as we now demonstrate, the orbit-averaged flux self-averages over the background fluctuations and the covariance $C_{\FE}(T_s,T_s')$ is suppressed.

The source is locally resolved, and the universal leading logarithm derived in Sec.~\ref{sec:large-Lambda} is local in time and on each worldline. It is therefore natural to separate the orbital average from the sum over bodies. For a fixed realization, the leading-logarithmic part of the total flux can be written as
\begin{equation}
\overline{\FE}^{\rm LL}(T_s)
\simeq
\frac{1}{T}\sum_A
\int_{T_s-T/2}^{T_s+T/2}\dd t\,
g_A^{\rm LL}(t)\rho(t,\bm x_A(t))\,,
\label{eq:leading-log-local-density-average}
\end{equation}
where $g_A^{\rm LL}(t)$ is the instantaneous leading-logarithmic flux per unit background density for body $A$.

Its connected covariance is therefore
\begin{align}
C_{\FE}^{\rm LL}(T_s,T_s')
\simeq{}&\frac{1}{T^2}\sum_{A,B}
\int_{T_s-T/2}^{T_s+T/2}\dd t
\int_{T_s'-T/2}^{T_s'+T/2}\dd t'
\nonumber\\
&\times
g_A^{\rm LL}(t)g_B^{\rm LL}(t')
C_\rho(t-t',\bm x_A(t)-\bm x_B(t')).
\label{eq:leading-log-flux-covariance}
\end{align}
This equation allows us to make the spatial self-averaging mechanism explicit and remains applicable to a general periodic orbit.

For a clean estimate, we specialize to a single body on a circular trajectory of radius $R$, for which $g^{\rm LL}$ is constant. The squared separation between two points on the orbit is
\begin{equation}
    |\bm x(t)-\bm x(t')|^2=4R^2\sin^2\!\left(\frac{\Omega\tau}{2}\right),
\end{equation}
where $\tau=t-t'$.

Inserting this result into the density correlation function, Eq.~\eqref{eq:maxwellian-density-correlation}, gives the following expression for the equal-time flux variance, $\avg{(\delta\overline{\FE}^{\rm LL})^2}=C_{\FE}^{\rm LL}(T_s,T_s)$:
\begin{align}
\frac{\avg{(\delta\overline{\FE}^{\rm LL})^2}}
{\avg{\overline{\FE}^{\rm LL}}^2}
\simeq{}&\frac{1}{T^2}\int_0^T\dd t\int_0^T\dd t'
\frac{1}{[1+(\tau/t_{\rm coh})^2]^{3/2}}
\nonumber\\
&\times\exp\!\left[
-4\frac{R^{2}}{\lambda_{\star}^{2}}\frac{\sin^2(\Omega\tau/2)}
{1+(\tau/t_{\rm coh})^2}
\right].
\label{eq:circular-many-patch-variance}
\end{align}
In both cases 6 and 7, $R\gg\lambda_\star$, and the integral in Eq.~\eqref{eq:circular-many-patch-variance} is localized near the coincidence saddle $t=t'$.  Writing $\tau=t-t'$, the width of this region is
\begin{equation}
t_{\rm cross}\sim\frac{\lambda_\star}{v}\ll T\,,
\end{equation}
so we can perform the integral, to leading order, by a saddle point approximation.  We obtain
\begin{align}
\frac{\avg{(\delta\overline{\FE}^{\rm LL})^2}}
{\avg{\overline{\FE}^{\rm LL}}^2}
\simeq
\sqrt{\pi}\frac{\lambda_\star}{vT}
=
\frac{\lambda_\star}{2\sqrt{\pi}R}.
\label{eq:coherent-many-patch-suppression}
\end{align}
Thus the rms fluctuation is suppressed relative to the mean value
\begin{equation}
\frac{\delta\overline{\FE}^{\rm LL}_{\rm rms}}
{\avg{\overline{\FE}^{\rm LL}}}
\simeq
\left(\frac{\lambda_\star}{2\sqrt{\pi}R}\right)^{1/2}\,.
\end{equation}
This is the expected $1/\sqrt{N}$ suppression characteristic of self-averaging, where parametrically $N\sim R/\lambda_\star$ is the number of distinct background patches traversed in one orbit.

We emphasize that this result applies to the universal leading-logarithmic contribution, whose locality in time was established in Sec.~\ref{sec:large-Lambda}.  The nonlogarithmic order-unity contribution can retain sensitivity to the full orbital history, and here we have not evaluated the covariance for these contributions.

\subsection{Fluctuations in kinetic-response regimes}
\label{subsec:kinetic-flux-fluctuations}

For each of the kinetic-response cases, $T/t_{\rm coh}\gg1$, so the typical temporal beat phase accumulates many cycles during one orbit. Cases 1 and 5 also sample many spatial patches. These rapidly varying phases suppress generic off-diagonal contributions to the orbit-averaged kernel. Unlike in the coherent-response cases 6 and 7, however, there is no local-density factorization from which the fluctuation relative to the mean can be determined. The following expressions make this distinction explicit.

At equal times, the reality of the orbit-averaged flux implies
\begin{equation}
\mathcal K_{T_s}(\bm k',\bm k)
=
\mathcal K_{T_s}(\bm k,\bm k')^*.
\end{equation}
Equations~\eqref{eq:kernel-mean-flux} and~\eqref{eq:kernel-flux-covariance} therefore give
\begin{equation}
\frac{C_{\FE}(T_s,T_s)}
{\left|\avg{\overline{\FE}}\right|^2}
=
\frac{
\displaystyle
\int_{\bm k,\bm k'}
P_\psi(k)P_\psi(k')
\left|\mathcal K_{T_s}(\bm k,\bm k')\right|^2
}{
\displaystyle
\left|
\int_{\bm k}
P_\psi(k)\mathcal K_{T_s}(\bm k,\bm k)
\right|^2
}.
\label{eq:kinetic-relative-fluctuation-ratio}
\end{equation}

Rapid beat phases can suppress the off-diagonal kernel appearing in the numerator. The denominator, however, depends only on the diagonal response and is not controlled by these beat phases. Moreover, its integrand is not sign definite. This is already apparent in Eq.~\eqref{eq:mean-flux-generalcase}: the factor $n\Omega$ changes sign, and the Doppler shift can place both positive- and negative-$n$ harmonics on shell. Different velocity and directional sectors can therefore contribute with opposite signs and partially cancel in the mean. Consequently, suppression of the off-diagonal kernel does not by itself imply $C_{\FE}(T_s,T_s)/|\avg{\overline{\FE}}|^2\ll1$.

An explicit determination of the relative fluctuation requires evaluating the full kernel in each kinetic regime. Cases 1 and 5 are both local and resolve the relative velocity of the environmental modes, whereas case 2 retains a multipolar source and can probe the kinetic--coherent crossover near $p\sim k_\star$. We do not perform these evaluations here. In all three cases, Eq.~\eqref{eq:kernel-flux-covariance}, together with the exact kernel~\eqref{eq:explicit-harmonic-flux-kernel}, provides the appropriate starting point for future work.

% ==================================================================
\section{Discussion and conclusions}
\label{sec:conclusion}

In this work, we have studied the conservative and dissipative forces on a Newtonian binary embedded in a nonrelativistic scalar medium, allowing the background field to be either coherent or stochastic. Starting from general expressions for the instantaneous force and orbit-averaged energy flux, we identified controlled limits in which the binary is resolved locally or only through its multipole moments, and in which the motion of the environment must be retained or can be treated coherently. Multipolar dissipation, local Coulomb drag, and realization-dependent fluctuations thereby emerge as different limits of the same wake-induced force.

For a coherent background, the conservative and dissipative sectors expose distinct pieces of the physics. The infrared divergence produced by the static mass monopole signals the failure of perturbation theory about a spatially constant scalar state. The gravitational Bohr scale $a_G$ identifies where this homogeneous expansion breaks down and must be replaced by a globally consistent, Coulomb-distorted background. The orbit-induced dissipative force is nevertheless controlled in the long-wavelength regime, where the scalar resolves the complete binary. The mass dipole-moment vanishes in the center-of-mass frame, and the leading flux is quadrupolar, scales as $\zeta^{5/2}$, and has the explicit eccentricity dependence derived above. In the opposite limit, a broad region of short-wavelength modes probes locally straight segments of the individual worldlines and generates the Coulomb logarithm $\log\sqrt{\zeta}$. The ratio of scales in the logarithm follows from orbital curvature and the on-shell scalar kinematics rather than being imposed externally. This universality applies to the logarithmic contribution; the accompanying terms of order $\zeta^0$ remain sensitive to the global orbit and can contain both conservative and dissipative components.

For a stochastic background, three logically independent questions were distinguished. The source can be multipolar or locally resolved, the environmental Green's function can reduce to its coherent limit or retain the full velocity distribution, and the orbit can either self-average over the random background or remain sensitive to a particular realization. The eight scale hierarchies identified in this work organize these possibilities. 

For the random-wave ensemble, the mean friction force depends only on the momentum-diagonal environmental response function, whereas its fluctuations retain interference between distinct background modes. We derived the corresponding connected two-time correlation function and evaluated it explicitly in the coherent regimes for a Maxwellian background distribution. When a locally resolved orbit crosses many patches, the universal leading-logarithmic contribution self-averages, with relative rms fluctuations suppressed as $(\lambda_\star/R)^{1/2}$. By contrast, a binary contained within one quasi-static patch inherits order-unity density fluctuations but does not self-average over a single orbit. This may have implications for the  accuracy of waveform modelling, however we leave such a study for future work. 

A natural next step is to apply these local force and covariance results to a self-consistent scalar environment which surrounds a compact object, and to propagate their effects through an evolving inspiral. Such an environment will generally have a spatially varying density and velocity distribution supported by the central gravitational potential, rather than the translation-invariant statistics assumed here. Extending the calculation to relativistic motion and curved-spacetime propagation would provide the corresponding framework for compact binaries in the strong-field regime.

A complete stochastic description must also combine the induced-wake fluctuations studied in this work with the direct gravitational force of the pre-existing interference pattern. Statistical correlations between the two forces is likely to occur, and this is presently unexplored. 

Another natural arena in which to extend the techniques developed here is cold dark matter. In this case, we only expect the analog of the  `hard' region discussed in \cref{sec:large-Lambda} to contribute, for which the presence of the primary should not affect the leading contribution. Indeed, in~\cite{kim2007dynamical} the IR cutoff is found to be approximately $2R$ from numerical fitting. However, in a realistic scenario the medium surrounding the binary is itself bound to the primary, rather than being homogeneous. In this case, the IR cutoff can be modified and may depend on the mass ratio of the binary. In recent work~\cite{Vicente:2025gsg} the IR cutoff is reasonably assumed to be the Hill radius. However, a full binary calculation will be necessary to derive the appropriate cutoff for a generic bound orbit, and the approach taken in this work may prove to be helpful for this purpose. Additionally, the EFT (and generalized-self-force) techniques may also be expanded to incorporate other effects like halo feedback~\cite{kavanagh2020detecting} within a common framework.

A central lesson of this work is that dynamical friction in a wave medium is not specified by an ambient density and a Coulomb logarithm alone. One must also determine which structures of the source and the medium are resolved, and whether the orbit samples enough of the background for an ensemble description to be representative. Depending on this hierarchy, the same scalar medium can produce a nearly deterministic orbital drift or a force with order-unity realization-dependent variations.

\section{Acknowledgements}
SM thanks Sumanta Chakraborty for helpful discussions on aspects of DF and SFDM. SM also wants to acknowledge the warm hospitality of IACS, Kolkata, where part of the work has been done. J.W.-G. thanks Ira Rothstein and Beka Modrekiladze for extensive discussions and collaboration on related work. J.W.-G. is supported by the US Department of Energy grant DE-SC001011, and by a President’s Postdoctoral Fellowship at CMU.
% ==================================================================
\appendix

\section{Diagrammatic derivation and master formulas}
\label{app:feynmandiagrams}

\subsection{Action, Fourier conventions, and binary source}
\label{app:action-conventions}

The nonrelativistic scalar and Newtonian potential are described by
\begin{equation}
S=\int\dd t\,\dd^3x
\left[
\ii\psi^*\partial_t\psi
-\frac{|\bm\nabla\psi|^2}{2\mu}
-\frac{(\bm\nabla\Phi)^2}{8\pi G}
-\mu|\psi|^2\Phi
-\rhoB\Phi
\right].
\label{eq:nonrelaction}
\end{equation}
The binary density is
\begin{equation}
\rhoB(t,\bm x)
=\sum_{A=1}^2m_A\delta^3(\bm x-\bm x_A(t)).
\label{eq:binary-density-position}
\end{equation}
We use
\begin{equation}
f(x)=\int_k e^{-\ii k\cdot x}f(k),
\qquad
k\cdot x=k^0t-\bm k\cdot\bm x,
\label{eq:fourier-convention}
\end{equation}
and
\begin{equation}
\int_k\equiv\int\frac{\dd k^0\,\dd^3k}{(2\pi)^4},\qquad\int_{\bm k}\equiv\int\frac{\dd^3k}{(2\pi)^3}.
\end{equation}
For later convenience define the time-local source insertion
\begin{equation}
\rhoB(p;t)
\equiv
\sum_{A=1}^2m_A
 e^{\ii p^0t-\ii\bm p\cdot\bm x_A(t)}.
\label{eq:binary-source-insertion}
\end{equation}
The first argument of $\chi(p,q)$ below is always the source-side Newtonian momentum $p^\mu$; the second is the receiver-side Newtonian momentum $q^\mu$. The internal scalar momentum is $k^\mu$. When translation invariance enforces $p=q$, we write $p=q=(\omega,\bm p)$.

We split
\begin{equation}
\psi=\psi_0+\delta\psi,
\end{equation}
where $\psi_0$ is a free background solution.  Expanding Eq.~\eqref{eq:nonrelaction} to linear order in $\delta\psi$ gives the mixing interaction
\begin{equation}
S_{\rm mix}
=-\mu\int\dd^4x\,\Phi
\left(\psi_0^*\delta\psi+\psi_0\delta\psi^*\right).
\label{eq:mixing-action}
\end{equation}

The source vertex is $-\ii m_A e^{\ii p\cdot x_{A}(t)}$; the receiving force vertex supplies $m_Aq^i e^{-\ii q\cdot x_A(t)}$.  These rules are most cleanly justified in the closed-time-path formalism, which fixes the retarded prescription and the response vertex.  At the classical level, the diagrammatic calculation is equivalent to solving the linearized equations of motion with retarded boundary conditions.

We use dashed lines for the background worldlines and dotted lines for background scalar fields. Solid scalar lines denote propagating scalar perturbations. Arrows on solid scalar lines indicate scalar-number flow, which we conventionally take to point from $\psi$ to $\psi^\ast$. Momentum labels are written separately and do not indicate the direction of scalar-number flow. The worldline index is $A=1,2$; the complete binary response is obtained by summing over the appropriate worldline insertions.

The worldline source and receiving vertices are
\begin{subequations}
\label{eq:worldline-feynman-rules}
\begin{align}
\begin{tikzpicture}[
    baseline=(current bounding box.center),
    x=0.85cm,
    y=0.85cm
]
  \draw[worldline] (-1.15,0) -- (1.15,0);
  \node[vertexdot] (v) at (0,0) {};
  \draw[graviton] (v) -- (0,1.05);
  % momentum p flowing away from the source
  \draw[->] (0.18,0.28) -- (0.18,0.78);
  \node[right=2pt] at (0.18,0.53) {$p$};
  \node[below=3pt] at (v) {$A$};
\end{tikzpicture}
&=
-\ii m_A e^{\ii p\cdot x_{A}(t)}\,,
\label{eq:worldline-source-rule}
\\[1.0em]
\begin{tikzpicture}[
    baseline=(current bounding box.center),
    x=0.85cm,
    y=0.85cm
]
  \draw[worldline] (-1.15,0) -- (0,0);
  \draw[extleg] (0,0) -- (1.15,0);
  \node[vertexdot] (v) at (0,0) {};
  \draw[graviton] (v) -- (0,1.05);
  % momentum q flowing into the receiver
  \draw[->] (0.18,0.78) -- (0.18,0.28);
  \node[right=2pt] at (0.18,0.53) {$q$};
  \node[below=3pt] at (v) {$A$};
  %\node[above=2pt] at (0.75,0) {$i$};
\end{tikzpicture}
&=
-\ii m_A\,(\ii q_i)e^{-\ii q\cdot x_{A}(t)}\,.
\label{eq:worldline-receiver-rule}
\end{align}
\end{subequations}

%%%%%%%%%%%%%%%%%%%%%%%%%%%%%%%%%%%%%%%%%%%%%%%%%%%%%%%%%%%%%%%%%%

The scalar--graviton mixing vertices are 
\begin{subequations}
\label{eq:scalar-mixing-rules}
\begin{align}
\begin{tikzpicture}[
    baseline=(current bounding box.center),
    x=0.90cm,
    y=0.90cm
]
  \node[vertexdot] (v) at (0,0) {};
%
  % Newtonian line carrying p into the vertex
  \draw[graviton] (v) -- (0,-1.0);
  \draw[->] (0.18,-0.78) -- (0.18,-0.28);
  \node[right=2pt] at (0.18,-0.53) {$p$};
%
  % background scalar carrying k-p
  \draw[bgscalar] (-1.05,0.75) -- (v);
  \node[above left=1pt] at (-1.0,0.20) {$k-p$};
%
  % internal scalar carrying k
  \draw[scalarflowrev] (v) -- (1.25,0);
  \node[above=3pt] at (0.82,0.25) {$k$};
  % momentum arrow on scalar line
  \draw[->] (0.5, 0.3) -- (1.0,0.3);
\end{tikzpicture}
&=
  -\ii\mu\,\widetilde{\psi}_0^{\,*}(k-p),
\label{eq:mixing-psi-rule}
\\[1.2em]
\begin{tikzpicture}[
    baseline=(current bounding box.center),
    x=0.90cm,
    y=0.90cm
]
  \node[vertexdot] (v) at (0,0) {};
%
  % Newtonian line carrying p into the vertex
  \draw[graviton] (v) -- (0,-1.0);
  \draw[->] (0.18,-0.78) -- (0.18,-0.28);
  \node[right=2pt] at (0.18,-0.53) {$p$};
%
  % background scalar carrying k-p
  \draw[bgscalar] (-1.05,0.75) -- (v);
  \node[above left=1pt] at (-1.0,0.20) {$k-p$};
%
  % internal scalar carrying k
  \draw[scalarflow] (v) -- (1.25,0);
  \node[above=3pt] at (0.82,0.25) {$k$};
  % momentum arrow on scalar line
  \draw[->] (0.5, 0.3) -- (1.0,0.3);
\end{tikzpicture}
&=
  -\ii\mu\,\widetilde{\psi}_0(k-p).
\label{eq:mixing-psistar-rule}
\end{align}
\end{subequations}

%%%%%%%%%%%%%%%%%%%%%%%%%%%%%%%%%%%%%%%%%%%%%%%%%%%%%%%%%%%%%%%%%%

The two oriented retarded scalar propagators are
\begin{subequations}
\label{eq:scalar-propagator-rules}
\begin{align}
\begin{tikzpicture}[
    baseline=(current bounding box.center),
    x=0.95cm,
    y=0.95cm
]
  \draw[scalarflow] (-1.35,0) -- (1.35,0);
  \node[above=3pt] at (0,0) {$k$};
\end{tikzpicture}
&=
G^R_{\delta\psi\delta\psi^\ast}(k)
=
-\frac{\ii}{k^0-E_{\bm k}+\ii0},
\label{eq:scalar-retarded-rule}
\\[1.0em]
\begin{tikzpicture}[
    baseline=(current bounding box.center),
    x=0.95cm,
    y=0.95cm
]
  \draw[scalarflowrev] (-1.35,0) -- (1.35,0);
  \node[above=3pt] at (0,0) {$k$};
\end{tikzpicture}
&=
G^R_{\delta\psi^\ast \delta\psi}(k)
=
-\frac{\ii}{k^0+E_{\bm k}+\ii0}.
\label{eq:scalar-conjugate-retarded-rule}
\end{align}
\end{subequations}

%%%%%%%%%%%%%%%%%%%%%%%%%%%%%%%%%%%%%%%%%%%%%%%%%%%%%%%%%%%%%%%%%%

The instantaneous Newtonian propagator is
\begin{equation}
\label{eq:potential-graviton-rule}
\begin{tikzpicture}[
    baseline=(current bounding box.center),
    x=0.90cm,
    y=0.90cm
]
  \draw[graviton] (-1.35,0) -- (1.35,0);
  \node[above=3pt] at (0,0) {$p$};
\end{tikzpicture}
=
-\frac{\ii\,4\pi G}{\bm p^2}.
\end{equation}

\begin{figure}[h!]
\centering
\begin{tikzpicture}[
    baseline=(current bounding box.center),
    x=0.95cm,
    y=0.95cm
]
  % bottom worldline vertices
  \node[vertexdot] (src) at (-1.9,0) {};
  \node[vertexdot] (rec) at (1.9,0) {};

  \draw[worldline] (-2.9,0) -- (-0.9,0);
  \draw[worldline] (0.9,0) -- (rec);
  \draw[extleg] (rec) -- (2.9,0);

  \node[below=3pt] at (src) {$B$};
  \node[below=3pt] at (rec) {$A$};
  %\node[above=2pt] at (2.75,0) {$i$};

  % upper mixing vertices
  %\node[vertexdot] (mixL) at (-1.15,1.35) {};
  %\node[vertexdot] (mixR) at ( 1.15,1.35) {};

  \node[vertexdot] (mixL) at (-0.95,1.30) {};
  \node[vertexdot] (mixR) at ( 0.95,1.30) {};

  % gravitons
  \draw[graviton] (src) -- (mixL);
  \draw[graviton] (mixR) -- (rec);

  % momentum arrows on gravitons
  \draw[->] (-2.25,0.35) -- (-1.65,1.05);
  \node[left=1pt] at (-1.93,0.72) {$p$};

  \draw[->] (1.65,1.05) -- (2.25,0.35);
  \node[right=1pt] at (1.93,0.72) {$q$};

  % dotted background insertions
  \draw[bgscalar] (-2.15,2.15) -- (mixL);
  \draw[bgscalar] ( 2.15,2.15) -- (mixR);

  \node[above left=1pt]  at (-1.80,1.48) {$k-p$};
  \node[above right=1pt] at ( 1.80,1.48) {$q-k$};

  % internal scalar line across the top
  \draw[scalarflow] (mixL) -- (mixR);
  \node[above=3pt] at (0,1.5) {$k$};
    % momentum arrow on scalar line
  \draw[->] (-0.5, 1.6) -- (0.5,1.60);
\end{tikzpicture}
\begin{tikzpicture}[
    baseline=(current bounding box.center),
    x=0.95cm,
    y=0.95cm
]
  % bottom worldline vertices
  \node[vertexdot] (src) at (-1.9,0) {};
  \node[vertexdot] (rec) at (1.9,0) {};

  \draw[worldline] (-2.9,0) -- (-0.9,0);
  \draw[worldline] (0.9,0) -- (rec);
  \draw[extleg] (rec) -- (2.9,0);

  \node[below=3pt] at (src) {$B$};
  \node[below=3pt] at (rec) {$A$};
  %\node[above=2pt] at (2.75,0) {$i$};

  % upper mixing vertices
  %\node[vertexdot] (mixL) at (-1.15,1.35) {};
  %\node[vertexdot] (mixR) at ( 1.15,1.35) {};

  \node[vertexdot] (mixL) at (-0.95,1.30) {};
  \node[vertexdot] (mixR) at ( 0.95,1.30) {};

  % gravitons
  \draw[graviton] (src) -- (mixL);
  \draw[graviton] (mixR) -- (rec);

  % momentum arrows on gravitons
  \draw[->] (-2.25,0.35) -- (-1.65,1.05);
  \node[left=1pt] at (-1.93,0.72) {$p$};

  \draw[->] (1.65,1.05) -- (2.25,0.35);
  \node[right=1pt] at (1.93,0.72) {$q$};

  % dotted background insertions
  \draw[bgscalar] (-2.15,2.15) -- (mixL);
  \draw[bgscalar] ( 2.15,2.15) -- (mixR);

  \node[above left=1pt]  at (-1.80,1.48) {$k-p$};
  \node[above right=1pt] at ( 1.80,1.48) {$q-k$};

  % internal scalar line across the top, reversed flow
  \draw[scalarflowrev] (mixL) -- (mixR);
  \node[above=3pt] at (0,1.50) {$k$};

  % momentum arrow on scalar line
  \draw[->] (-0.5, 1.6) -- (0.5,1.60);
\end{tikzpicture}
\caption{The two scalar orientations contributing to the dynamical friction force. Dashed horizontal lines denote background worldlines, dotted lines denote background scalar insertions, solid scalar lines denote propagating scalar perturbations, and their arrows denote scalar particle-number flow. Separate arrows indicate momentum routing.}
\label{fig:app_force_diagrams}
\end{figure}
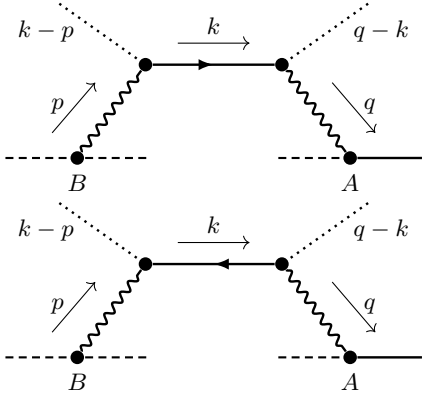

The two scalar orientations contributing to the dynamical-friction force on body $A$, sourced by body $B$, are shown in \cref{fig:app_force_diagrams}.
%%%%%%%%%%%%%%%%%%%%%%%%%%%%%%%%%%%%%%%%%%%%%%%%%%%%%%%%%%%%%%%%%%

\subsection{General response kernel and force}
\label{app:general-kernel}

Writing $\widetilde\psi_0(k)$ for the Fourier transform defined by Eq.~\eqref{eq:fourier-convention}, summing the two scalar orientations defines the momentum-nondiagonal density response
\begin{align}\label{eq:chi-general}
\scalebox{0.97}{$\displaystyle
\chi(p,q) =\mu^2\int_k\Bigg[
\frac{\widetilde\psi_0(k-p)\widetilde\psi_0^*(q-k)
}
{k^0-E_{\bm k}+\ii0}
-\frac{\widetilde\psi_0^*(k-p)\widetilde\psi_0(q-k)
}
{k^0+E_{\bm k}+\ii0}
\Bigg].
$}
\end{align}

For a generic realization, the background carries the momentum mismatch $q-p$, and $\chi(p,q)$ is not diagonal.

The instantaneous force on body $A$ follows from summing over both source bodies:
\begin{align}
&F_A^i(t) = \nonumber \\
&-\ii(4\pi G)^2m_A
\int\dd t'\int_{p,q}
 e^{-\ii q\cdot x_A(t)}
\rhoB(p;t')
\frac{q^i}{\bm p^2\bm q^2}
\chi(p,q).
\label{eq:forceinstant2general}
\end{align}
Causality is carried by the retarded denominators in Eq.~\eqref{eq:chi-general}. We will also be interested in the total orbit-averaged energy flux from the binary into the environment,
\begin{equation}
\overline{\FE}
=-\frac{1}{T}\sum_{A=1}^2
\int_0^T\dd t\,
\dot{\bm x}_A(t)\cdot\bm F_A(t).
\label{eq:flux-definition-appendix}
\end{equation}
Integrating by parts in $t$ and using periodicity gives
\begin{align}
&\overline{\FE}=\nonumber \\
&\frac{\ii(4\pi G)^2}{T}
\int_0^T\dd t\int\dd t'\int_{p,q}
\rhoB(-q;t)\rhoB(p;t')
\frac{q^0}{\bm p^2\bm q^2}
\chi(p,q).
\label{eq:orbitaveragedenergeneral}
\end{align}
Equations~\eqref{eq:forceinstant2general} and \eqref{eq:orbitaveragedenergeneral} are the general master formulas used in the stochastic sections.

\subsection{Constant coherent background}
\label{app:coherent-limit}

For a constant real background $\psi_0$, the mass density is
\begin{equation}
\rho_0=\mu \psi_{0}^2,
\end{equation}
Equation~\eqref{eq:chi-general} then becomes
\begin{equation}
\chi(p,q)
=(2\pi)^4\delta^4(p-q)\,
\rho_0\frac{\bm p^2}
{(\omega+\ii0)^2-E_{\bm p}^2},
\label{eq:coherent-chi}
\end{equation}
Translation invariance therefore diagonalizes and simplifies the force and flux integrals.

For a periodic binary, we can further simplify the general expression by defining the source harmonics
\begin{equation}
\rhoB(p;t)
=e^{\ii \omega t}\sum_{n=-\infty}^{\infty}
\Sorb_n(\bm p)e^{-\ii n\Omega t},
\label{eq:binary-source-harmonic-expansion}
\end{equation}
with
\begin{equation}
\Sorb_n(\bm p)
=\frac{1}{T}\int_0^T\dd t\,
 e^{\ii n\Omega t}
\left[
 m_1e^{-\ii\bm p\cdot\bm x_1(t)}
 +m_2e^{-\ii\bm p\cdot\bm x_2(t)}
\right].
\label{eq:binary-harmonic-source}
\end{equation}
The time and frequency integrals can then be performed explicitly, giving compact expressions in the frequency-harmonic basis. The instantaneous force on body $A$ is
\begin{align}
&F_A^i(t) =-\ii(4\pi G)^2\rho_0m_A
\sum_{n=-\infty}^{\infty}
\nonumber \\ 
&\times\intp
 e^{-\ii n\Omega t+\ii\bm p\cdot\bm x_A(t)}
\Sorb_n(\bm p)\frac{p^i}{\bm p^2}\frac{1}{(n\Omega+\ii0)^2-E_{\bm p}^2}.
\label{eq:coherentforceformula}
\end{align}
while the total orbit-averaged energy flux from the binary into the environment is
\begin{equation}
\overline{\FE}
=\pi(4\pi G)^2\rho_0
\sum_{n=1}^{\infty}\intp
\frac{|\Sorb_n(\bm p)|^2}{\bm p^2}
\delta\!\left(n\Omega-E_{\bm p}\right),
\label{eq:coherentfluxformula}
\end{equation}
Only the imaginary part of the response function $\chi(p,q)$ contributes to this dissipative quantity, producing the energy-conserving delta function. The result is manifestly nonnegative and includes all two-body interference terms. Equations~\eqref{eq:coherentforceformula} and~\eqref{eq:coherentfluxformula} are the general formulas used in the coherent-field sections.

\bibliography{references} 

\end{document}

%% file: tikz_styles.tex
\usetikzlibrary{
  arrows.meta,
  decorations.pathmorphing,
  decorations.markings,
  positioning
}

\tikzset{
  worldline/.style={
    dash pattern=on 3pt off 2pt,
    line width=0.8pt
  },
  graviton/.style={
    decorate,
    decoration={snake, amplitude=1.4pt, segment length=5pt},
    line width=0.9pt
  },
  bgscalar/.style={
    dotted,
    line width=0.9pt
  },
  scalarflow/.style={
    line width=0.9pt,
    postaction={decorate},
    decoration={
      markings,
      mark=at position 0.5 with {
        \arrow{Latex[length=2.0mm,width=1.5mm]}
      }
    }
  },
  scalarflowrev/.style={
    line width=0.9pt,
    postaction={decorate},
    decoration={
      markings,
      mark=at position 0.5 with {
        \arrowreversed{Latex[length=2.0mm,width=1.5mm]}
      }
    }
  },
  extleg/.style={
    line width=0.9pt
  },
  vertexdot/.style={
    circle,
    fill=black,
    inner sep=1.8pt
  }
}